\documentclass[useAMS,usenatbib,babel]{mn2e}
\usepackage[a4paper, margin=0.75in]{geometry}
\usepackage{multirow,balance,verbatim}
\usepackage[english,english]{babel}
\usepackage{amsmath,subfig}
\usepackage{amssymb,amsfonts,textcomp}
\usepackage{array,float}
\usepackage{supertabular}
\usepackage{hhline}
\usepackage{hyperref}
\usepackage[usenames]{color}
\usepackage[dvipsnames]{xcolor}
\hypersetup{dvips, colorlinks=true, linkcolor=black, citecolor=black, filecolor=black, urlcolor=black}
\usepackage[dvips]{graphicx}
\usepackage[squaren,Gray,cdot]{SIunits}
\usepackage{balance}
\def\gtrsim{\lower.5ex\hbox{$\; \buildrel > \over \sim \;$}}

\def \prd {{\it Phys. Rev. D}}

\def \apj {{\it ApJ}}
\def \apjl {{\it ApJ Let.}}
\def \apjs {{\it ApJ Sup.}}

\def \aap {{\it A\&A}}

\def \mnras {{\it MNRAS}}

\def \physrep{{\it Phys.~Rep.}}
\def \pasp{{\it PASP}}

\begin{document}

\author[Chisari et al.]{
  \parbox{\textwidth}{N. E. Chisari$^1$\thanks{elisa.chisari@physics.ox.ac.uk}, N. Koukoufilippas$^{2}$, A.~Jindal$^3$, S.~Peirani$^4$, R. S. Beckmann$^1$, S.~Codis$^3$, J.~Devriendt$^1$,  L. Miller$^1$,  Y. Dubois$^4$, C. Laigle$^1$, A. Slyz$^1$, C. Pichon$^{4,5}$}
\vspace*{6pt}\\
\noindent
$^{1}$Department of Physics, University of Oxford, Keble Road, Oxford, OX1 3RH,UK.\\
$^{2}$School of Physics and Astronomy, Cardiff University, The Parade, Cardiff, CF24 3AA, UK.\\
$^{3}$Canadian Institute for Theoretical Astrophysics, University of Toronto, 60 St. George Street, Toronto, ON M5S 3H8, Canada.\\
$^{4}$Institut d'Astrophysique de Paris, CNRS \& UPMC, UMR 7095, 98 bis Boulevard Arago, 75014, Paris, France.\\
$^{5}$Korea Institute of Advanced Studies (KIAS) 85 Hoegiro, Dongdaemun-gu, Seoul, 02455, Republic of Korea.
}

\date{Accepted 2017 August 2. Received 2017 August 2; in original form 2017 February 21.}

\title[Galaxy-halo alignments]{
Galaxy-halo alignments in the Horizon-AGN cosmological hydrodynamical simulation}

\maketitle

\begin{abstract}
  Intrinsic alignments of galaxies are a significant astrophysical systematic affecting cosmological constraints from weak gravitational lensing. Obtaining numerical predictions from hydrodynamical simulations of expected survey volumes is expensive, and a cheaper alternative relies on populating large dark matter-only simulations with accurate models of alignments calibrated on smaller hydrodynamical runs. This requires connecting the shapes and orientations of galaxies to those of dark matter haloes and to the large-scale structure. In this paper, we characterise galaxy-halo alignments in the Horizon-AGN cosmological hydrodynamical simulation. We compare the shapes and orientations of galaxies in the redshift range $0<z<3$ to those of their embedding dark matter haloes, and to the matching haloes of a twin dark-matter only run with identical initial conditions. We find that galaxy ellipticities in general cannot be predicted directly from halo ellipticities. The mean misalignment angle between the minor axis of a galaxy and its embedding halo is a function of halo mass, with residuals arising from the dependence of alignment on galaxy type, but not on environment. Haloes are much more strongly aligned among themselves than galaxies, and they decrease their alignment towards low redshift. Galaxy alignments compete with this effect, as galaxies tend to increase their alignment with haloes towards low redshift. We discuss the implications of these results for current halo models of intrinsic alignments and suggest several avenues for improvement. 
\end{abstract}


\section{Introduction}
\label{sec:intro}

Galaxy shapes present intrinsic correlations with the large-scale structure of the Universe known as ``intrinsic alignments''\citep{Croft00,Lee00,Heavens00,Catelan01,Crittenden01,Mackey02,Brown02,Hey++04,aubertetal04,Mandelbaum06,Joachimi11,Singh14,Singh15}. These correlations are the main astrophysical systematic affecting cosmological constraints from weak gravitational lensing across all scales. Failing to account for them can result in biased constraints on the equation of state of dark energy from future survey data \citep{Hirata04,Hirata07,Kirk12,Krause15}. Several groups have characterised intrinsic alignments in numerical hydrodynamical simulations \citep{Codis14,Tenneti14a,Tenneti15a,Tenneti15b,Velliscig15,Velliscig15b,Chisari15,Chisari16,Hilbert16}, in an effort to learn about the physical origin of these correlations and to create accurate models for incorporation into future survey data analysis pipelines.

However, cosmological hydrodynamical simulations cannot achieve the enormous volumes that will be probed by future surveys. State-of-the-art hydrodynamical simulations have typical volumes of $(100$ Mpc$/h)^3$ \citep{Dubois14,Vogelsberger14,Schaye15,Khandai15}, while galaxy surveys such as {\it Euclid}\footnote{\url{http://sci.esa.int/euclid}} or the Large Synoptic Survey Telescope (LSST\footnote{\url{https://www.lsst.org}}), will probe multi-Gpc$^3$ cosmological volumes. The cost of following the evolution of the gas and star formation with sufficient resolution is prohibitive in such volumes and thus another approach must be taken \citep{Kiessling15}. The alternative is to use dark matter-only (DMO) simulations, which only require following the effect of the gravitational force on the dark matter, and can thus achieve the required multi-Gpc volumes with typical $10^{10}$ M$_\odot$ dark matter mass resolution at a lower cost. To make predictions for the observables of astronomical surveys, DMO simulations must be populated with galaxies bearing some statistical relation with their embedding dark matter haloes \citep{Heymans06}. Haloes have been shown to be subject to alignments up to large scales \citep{Croft00,Heavens00,Jing02,Faltenbacher02,aubertetal04,Bailin05,Kasun05,Hopkins05,Smargon12}. At the very least, the prescribed relation should provide the misalignment angle between the galaxy and the halo, and the galaxy ellipticity, as a function of some halo properties. In addition, to go beyond the limit of resolution of current DMO simulations and model observable ``satellite'' galaxies, one must specify how to populate haloes with satellites including a prescription for some level of alignment according to a `halo model' \citep{Seljak00,Cooray02,Okumura09,Schneider10,Joachimi13a,Joachimi13b}. 

In this manuscript, we test the assumptions behind such a model for populating the dark matter haloes from DMO simulations with aligned galaxies, using the measured intrinsic alignments from the Horizon-AGN simulation\footnote{\url{http://www.horizon-simulation.org}} \citep{Dubois14,Kaviraj16} in the redshift range of interest of future weak lensing surveys, $0<z<3$. The shape, orientation of the minor axis and direction of the angular momentum of each galaxy is compared to that of its embedding halo, and to a corresponding halo in the twin DMO simulation, Horizon-DM, which was run from identical initial conditions. We also quantify the strength of halo alignments by computing the correlations of their three-dimensional shapes and orientations of their angular momenta with the large-scale structure. The availability of the twin DMO run further allows us to explore the impact of baryonic processes on the alignment of dark matter haloes.

Other groups have studied galaxy-halo misalignments in the EAGLE \citep{Velliscig15} and MassiveBlack-II \citep{Tenneti14a} cosmological hydrodynamical simulations. These simulations use a different numerical technique (smoothed-particle-hydrodynamics) and different choices of baryonic physics parameters than Horizon-AGN (which uses an adaptive-mesh-refinement method). Some discrepancies have been found between them \citep{Chisari16,Tenneti15b}, particularly in the angular momentum-alignment of low mass galaxies with the large-scale structure. It is thus important to provide intrinsic alignment halo models capable of reproducing the measurements of the different simulations, and to identify where those differences lie. The galaxy-halo alignment measurements presented here are the first to be obtained from a cosmological adaptive-mesh-refinement (AMR) simulation.

This manuscript is organised as follows. Section \ref{sec:sim} summarises the characteristics of the Horizon-AGN simulation and its twin DMO run, Horizon-DM. There, we also present the details of how galaxies and haloes are identified, correspondingly matched across simulations, their shapes measured and the cosmic web structure extracted. The alignment statistics are presented in Section \ref{sec:correl}, while we show we can safely neglect numerical systematics in appendix \ref{app:gridlock}. We present our results in Section \ref{sec:results} and discuss them in the context of other numerical works in Section \ref{sec:discuss}. Some further results which pertain to the discussion are presented in appendix \ref{app:projected}. Our conclusions are summarised in Section \ref{sec:conclusion}.

\section{The Horizon-AGN \& Horizon-DM simulations}
\label{sec:sim}

Horizon-AGN is a state-of-the-art cosmological hydrodynamical simulation of size $L= 100\,h^{-1}$ Mpc on each side, following the evolution of dark matter and baryons, and the formation of galaxies to the present. The simulation was performed using the AMR code {\sc ramses} \citep{teyssier02} modelling gas dynamics, cooling and heating \citep{sutherland&dopita93}, and various sub-grid processes. Because of the adaptive grid, the minimum cell size achieved is $1$ kpc constant in physical length. There are $1024^3$ dark matter particles, resulting in a minimum resolution of $\mathcal{M}_{DM}=8\times 10^7$ M$_\odot$.

Star formation is triggered according to a Poisson random process \citep{rasera&teyssier06,dubois&teyssier08winds} when the Hydrogen gas number density exceeds a threshold of $0.1$ H cm$^{-3}$. Star formation follows a Schmidt law with constant star formation efficiency, $\dot \rho_*=0.02\rho/t_{ff}$ \citep{kennicutt98,krumholz&tan07}, where $\rho$ is the density of the gas and $t_{ff}$, its local free-fall time. The resulting stellar mass resolution is $\mathcal{M}_*=2\times 10^6$ M$_\odot$. Feedback from stellar winds and type II supernovae is included using {\sc starburst}99 \citep{leithereretal99,leithereretal10}. The type Ia rate is taken from \citet{greggio&renzini83}. The model for feedback from active galactic nuclei (AGN) is described in \citet{duboisetal12agnmodel} and \citet{Dubois14}.

Horizon-AGN adopts a WMAP7 cosmology \citep{komatsuetal11}, with the following cosmological parameters: $\{\Omega_m,\Omega_b,\Omega_\Lambda,h,\sigma_8,n_s\}=\{0.272,0.045,0.728,0.704,0.81,0.967\}$, where $\Omega_m$ is the total matter density, $\Omega_b$ is the baryon density, $\Omega_\Lambda$ is the dark energy density, $h=H_0/(100$ km s$^{-1}$ Mpc$^{-1})$ is normalized value of the Hubble constant today, $\sigma_8$ is the amplitude of the power spectrum of density fluctuations averaged on spheres of $8 h^{-1}\,\rm Mpc$ radius today and $n_s$ is the power-law index of the primordial power spectrum.

Horizon-DM has the same volume and identical initial conditions as Horizon-AGN, but it is a purely DMO run, without baryons. Comparison between Horizon-AGN and Horizon-DM allows us to constrain the impact of certain baryonic processes in the large-scale distribution of dark matter.

\subsection{Galaxy \& halo catalogues}
\label{sec:cats}

Galaxies are identified in the Horizon-AGN using the {\sc AdaptaHOP} algorithm \citep{aubertetal04}. The algorithm directly uses the distribution of stellar particles to select relevant over-densities. The criterion is the following: a structure is identified if the local density computed from the twenty nearest neighbours exceeds $178$ times the cosmological average density, and if it contains more than $50$ particles. The total stellar mass of a galaxy is given by the sum of the masses of its stellar particles.

Each stellar particle has a velocity that can be decomposed into cylindrical coordinate components: $v_r$, $v_\theta$ and $v_z$. The dynamical properties of a galaxy are quantified by the $V/\sigma_V$ parameter. This is the ratio between the mean rotational velocity of a galaxy, $V=\langle v_\theta\rangle$, and the average velocity dispersion, $\sigma_V^2 = (\sigma_r^2+\sigma_\theta^2+\sigma_z^2)/3$. We will refer to disc galaxies as those with $V/\sigma_V>0.6$ and to ellipticals as those with $V/\sigma_V\leq 0.6$, similarly to \citet{Dubois14}; small variations in the choice of this threshold do not impact our results.

\citet{Kaviraj16} compared luminosity and mass functions, the star formation main sequence and rest-frame UV-optical-near infrared colours, and the cosmic star formation history in Horizon-AGN to observations in the redshift range $0<z<6$. They found an overall good agreement between the predicted statistical quantities from the simulated universe with observations, in particular when uncertainties and systematics on both sides (observations and simulation) are taken into account. However, they highlight also some points of tension, specifically the overproduction of low mass galaxies and the under-production of massive galaxies at high redshift. The same problem has been identified in the Illustris simulation \citep[figure 2]{Genel14} and to less extent in the EAGLE simulation \citep[figure 3]{Furlong14} and this discrepancy with observations is specifically tied to the choice of subgrid recipes. \citet{Dubois16} studied the morphological diversity of galaxies in Horizon-AGN, comparing it to observations and to a twin simulation run lacking AGN feedback (Horizon-noAGN). They found that AGN feedback is determinant in triggering a morphological transformation of massive galaxies into ellipticals, and reproducing the observed fraction of ellipticals at low redshift \citep{Conselice06}. Without AGN feedback, a disc is typically rebuilt due to continued gas accretion and star formation, and despite the action of galaxy mergers. The presence of AGN feedback improves the match with the observed stellar-halo mass relation and the size-mass relation of galaxies. Nevertheless, low mass galaxies at high redshift suffer from too much star formation, presumably because stellar feedback is not strong enough; while at low redshift, low mass galaxies are more passive than in observations since they have completely used the gas at their disposal. Horizon-AGN was {\it not} tuned to match morphological properties of galaxies; the only tuning concerns the black hole-galaxy mass relation \citep{duboisetal12agnmodel,Volonteri16}. 

We present the results of this work on galaxy-halo alignments as a function of several galaxy properties. Galaxies are divided into three stellar mass bins: $M\leq 10^{9.5}$ M$_\odot$, $10^{9.5}$ M$_\odot$ $< M \leq 10^{10.5}$ M$_\odot$ and $M> 10^{10.5}$ M$_\odot$. The median halo mass in these bins is: $\langle M_h\rangle = 10^{10.8}$ M$_\odot$, $\langle M_h\rangle = 10^{11.2}$ M$_\odot$ and $\langle M_h\rangle = 10^{12}$ M$_\odot$, respectively. The specific choice of mass bins is not particularly important, as it is only intended to provide a qualitative insight into alignment trends. We explore some of these trends in more detail as a function of halo mass when necessary. We also study the residual dependence of alignments on dynamical properties of galaxies and on environment.

Similarly to galaxies, haloes are identified by applying the {\sc AdaptaHOP} algorithm \citep{aubertetal04}. The centre of the halo is temporarily defined as the ``densest particle'' in the halo, where the density is computed from the $20$ nearest neighbours. In a subsequent step, we draw a sphere of the size of the virial radius around it and implement a shrinking sphere method \citep{Power03} to recursively find the centre of the halo. In each iteration, the radius of the halo is reduced by $10\%$. The search is stopped when a sphere $3$ times larger than our spatial resolution is reached, and the final centre is the densest particle in that sphere. In the final catalogue, we only consider structures with more than $100$ particles and with more than $80$ times the average density of the box. The mass of the halo is defined as the sum of the masses of the member dark matter particles. Notice that {\sc AdaptaHOP} identifies a hierarchy of haloes and subhaloes by using information on the local density and the connectivity between the particles. Subhaloes that satisfy our criteria set above are also included in our sample.

\subsection{Shapes \& spins}
\label{sec:shapedef}

Galaxy and halo shapes are modelled by three-dimensional ellipsoids, whose axes point in the direction of the eigenvectors of the simple inertia tensor ($SIT$),
\begin{equation}
  I_{ij} = \frac{1}{M} \sum_{n} m_{(n)} x^{(n)}_i x^{(n)}_j,
  \label{eq:sit}
\end{equation}
where $i,j=\{1,2,3\}$ correspond to the axes of the simulation box, $m_{(n)}$ is the mass of the $n$-th particle and $M$ is the total mass of the structure. In the case of galaxies, the positions of the particles are measured with respect to the centre of mass, but we have verified that using the densest particle as center does not impact our conclusions. For haloes, we adopt the definition of centre as described in Section \ref{sec:cats}. The eigenvalues of the inertia tensor are related to the length of each axis of the ellipsoid by: $a=\sqrt{\lambda_a}$, $b=\sqrt{\lambda_b}$ and $c=\sqrt{\lambda_c}$ from largest to smallest axis, respectively. The axis ratios are $q=b/a$ and $s=c/a$.

An alternative proxy to determine the shape is the reduced inertia tensor ($RIT$),
\begin{equation}
  I^{R}_{ij} = \frac{1}{M}\sum_{n} m_{(n)} \frac{x^{(n)}_i x^{(n)}_j}{r_{(n)}^2},
  \label{eq:rit}
\end{equation}
where each particle is weighted by the inverse square distance to the centre. This procedure up-weights stellar or dark matter particles closer to the centre, roughly mimicking a luminosity-weighting scheme, which would be closer to observational estimates of galaxy shapes \citep{Tenneti15a}.

The shapes of structures with small number of particles can be biased due to insufficient resolution. Following the criteria set in our previous work \citep{Chisari15} and for consistency, we restrict our study to galaxies with more than $300$ stellar particles (resp. haloes and dark matter particles). This effectively restricts the galaxy population to those with stellar masses larger than $10^{9}$ M$_\odot$.

We also define the spin of haloes (resp. galaxies) as the intrinsic angular momentum of their dark matter (resp. stellar) particles
\begin{equation}
{\bf l}=\sum_{n}m^{( n)} \mathbf{x}^{(n)} \times \mathbf{v}^{(n)} \, ,
\end{equation}
where $\mathbf{v}^{(n)}$ is the velocity of each particle relative to the centre of mass. For this work, we are in general only interested in the direction of the angular momentum, and not in its magnitude.

\subsection{Cosmic web extraction}
\label{sec:cosmicweb}

In order to investigate the effect of the large-scale environment on the shape of galaxies and haloes, we extract the filaments of the cosmic web using a topological algorithm, DISPERSE \citep{2011MNRAS.414..350S,2011MNRAS.414..384S}. This code is a generalisation of the ``skeleton'' picture \citep{sousbie09} for discrete tracers. It uses discrete topology to extract the so-called Morse-Smale complex associated with the distribution of galaxies.
Filaments are then defined as the gradient lines joining maxima (i.e the nodes of the cosmic web) and saddle points of signature $+--$ (two negative eigenvalues). In this context, topological persistence, defined as the absolute difference between the value of the density field at the node and at its linked saddle point, is used to keep only the most prominent filaments and is a way to filter out noisy structures. 
In practice, we extract the persistent cosmic web from a Delaunay tessellation of the galaxy distribution, for different thresholds of persistence from $N_\sigma=3$ to $7$ \footnote{This procedure removes any persistence pair -- composed of a maximum and a saddle point -- with probability less than $N_\sigma$ times the dispersion to appear in a Gaussian random field.}. In section \ref{sec:resid} we present results corresponding to $N_\sigma=7$ persistence, but we have verified that reducing this threshold does not affect our conclusions.

\subsection{Matching between simulations}
\label{sec:matching}

\begin{figure}
  \centering
  \includegraphics[width=0.47\textwidth]{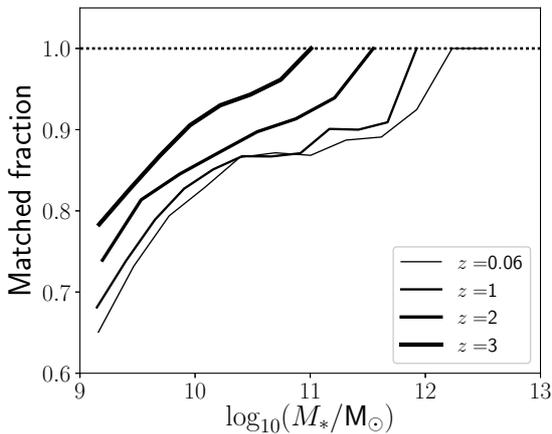}
  \caption{The fraction of galaxies with more than $300$ stellar particles that have been matched to haloes in the Horizon-AGN simulation and in the Horizon-DM simulation at different redshifts, as a function of stellar mass. The dotted line indicates the optimal case in which no galaxies would be lost through the matching between twin runs.}
  \label{fig:matcheff}
\end{figure}
\begin{figure}
  \centering
  \includegraphics[width=0.47\textwidth]{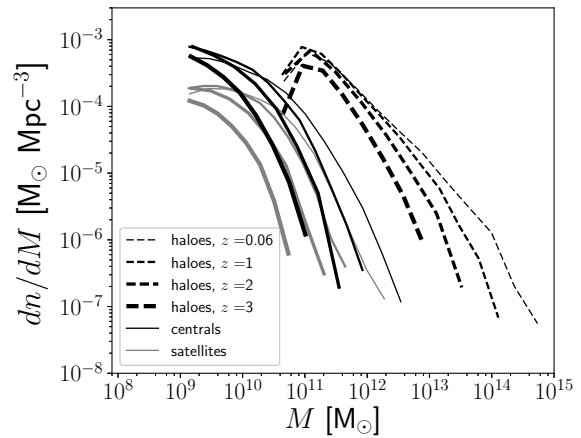}
  \caption{This figure presents the mass function of central galaxies with more than $300$ stellar particles and matched to haloes in both Horizon-AGN and Horizon-DM simulation (solid black), and of the corresponding haloes in Horizon-AGN (dashed black). The $x$ axis represents the stellar or dark matter mass for galaxies or haloes, respectively. The solid gray curves represent the mass function of satellite galaxies. Increasing thickness corresponds to increasing redshift. The central halo mass function of Horizon-DM would be shifted ($\sim 40\%$) to higher masses than the Horizon-AGN haloes, and for clarity it is not shown here.}
  \label{fig:massfunc}
\end{figure}

At $z=0.06$, and before cross-matching, there are $12 \times 10^4$ galaxies and $31 \times 10^4$ dark matter haloes and subhaloes in the Horizon-AGN simulation, and $37\times 10^4$ haloes and subhaloes in the Horizon-DM simulation at the same redshift. Because the baryonic and DMO simulations are run with identical initial conditions, it is possible to cross-match the haloes in the two simulations \citep{Peirani16,Beckmann17}. To identify the halo counterparts of Horizon-AGN in Horizon-DM, we look for haloes that have at least $50\%$ of their dark matter particles in common. (Particles are assigned an index at the initial conditions and this index is stored throughout the simulation.) This gives rise to multiple counterparts, as a given dark matter halo in Horizon-DM can be matched to several Horizon-AGN haloes due to the presence of substructure. Among them, we pick the matched pair which is most similar in mass. This choice would effectively remove most of higher order objects in the substructure hierarchy. To avoid this bias, we repeat the matching in the other direction, looking for counterparts to the Horizon-DM haloes. The efficiency of this matching is not only determined by the choice of algorithm, but also limited by the impact of baryonic effects on the haloes in Horizon-AGN. 

To match galaxies to their haloes in Horizon-AGN, the procedure is different. In this case, we look for the galaxies that are located within $10\%$ of the virial radius of the dark matter halo in Horizon-AGN, and choose the most massive among them as the match. Through the halo matching, stellar haloes can also be indirectly matched to haloes in Horizon-DM. The efficiency of this matching can be observed in Figure \ref{fig:matcheff}. The solid curves show the fraction of galaxies with more than $300$ stellar particles which are matched to a dark matter halo in Horizon-AGN and Horizon-DM at different redshifts. Among the galaxies matched, we will define a sample of ``satellite'' galaxies as those hosted by Horizon-DM subhaloes; while ``centrals'' will be those associated with Horizon-DM haloes. We purposely use the Horizon-DM hierarchy to define the satellites, as this is the information accessible by larger-volume DMO simulations.

The number of galaxies matched to Horizon-AGN haloes and Horizon-DM haloes at each redshift is: $\{65059,70250,56792,30689\}$ for $z=\{0,1,2,3\}$, respectively. The percentage of central galaxies at each redshift is $\{72,76,79,82\}\%$, with the remaining fraction being classified as satellites. The mass functions of matched haloes, centrals and satellites at each redshift are shown in Figure \ref{fig:massfunc}. Note that the specific parameters of the matching, i.e. the percentage of overlapping DM particles and the maximum separation between galaxy and halo centre, are chosen specifically for this application. Other works using the Horizon suite of simulations \citep{Peirani16,Beckmann17} have adopted different thresholds more suitable to their investigations.

\begin{figure}
  \centering
  \includegraphics[width=0.47\textwidth]{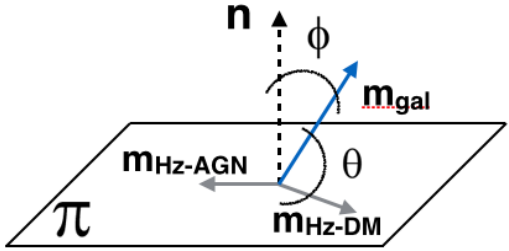}
  \caption{A sketch of the relation between the minor axis of a galaxy (${\bf m}_{\rm gal}$), the minor axis of the host Horizon-AGN halo (${\bf m}_{\rm Hz-AGN}$) and that of the Horizon-DM halo (${\bf m}_{\rm Hz-DM}$). The minor axes of the haloes define a plane $\pi$, with normal unit vector ${\bf n}$. The angle between the minor axis of a galaxy and the normal vector is $\phi$; and we will generally refer to the angle between the galaxy's minor axis and either halo's minor axis as $\theta$.}
  \label{fig:cartoon}
\end{figure}

\section{Quantifying alignments}
\label{sec:correl}

We define the (misalignment) angle, $\theta$, between the orientation of the axis of a galaxy and a halo, either in Horizon-AGN or in Horizon-DM. For two parallel or anti-parallel vectors, $\theta=0$; while for two perpendicular vectors, $\theta=90^\circ$. Unless explicitly noted, we are typically interested in the angle between the minor axes of the ellipsoids. We have found this is a more robust measurement of galaxy alignment than the major axes \citep{Chisari15}. In Section \ref{sec:results}, we quantify the level of alignment between galaxies and haloes by estimating the probability density function (PDF) of $\theta$ and comparing it to the PDF of angles distributed randomly, in which case the mean expectation is $57^\circ$. 

The minor axis of the Horizon-AGN halo, and that of the Horizon-DM halo, define a plane $\pi$ characterised by a normal unit vector, ${\bf n}$, as shown in Figure \ref{fig:cartoon}. The minor axis of a galaxy does not necessarily live on the plane $\pi$, hence we define $\phi$ as the angle between the galaxy minor axis and this plane. In practice, although this quantity provides additional physical insights, it is not necessary for populating DMO simulations; to that end, we only need the distribution of angles between the galaxy and the Horizon-DM halo minor axes. We will alternatively consider the direction of the intrinsic angular momentum of a galaxy or halo (``spin") as a proxy of galaxy orientation. As we have shown in our previous work, the spin and the minor axis of a galaxy tend to coincide for disc galaxies in the simulation; while they are less correlated for ellipticals \citep{Chisari15,Chisari16}.

We quantify alignments between haloes and galaxies through the position-orientation correlation, constructed by obtaining
\begin{equation}
  \eta(r) = \langle \cos^2\beta \rangle - 1/3,
  \label{eq:eta}
\end{equation}
where $\beta$ is the angle between the minor axis and the separation vector of two haloes (or galaxies), separated by comoving distance $r$. A negative $\eta(r)$ indicates a preferential perpendicular alignment of the minor axis with respect to the separation vector. On the other hand, positive $\eta(r)$ corresponds to the minor axis and the separation vector pointing preferentially in the same direction. A null result corresponds to no alignment.

We are also interested in the average shapes of galaxies, as given by the axis ratios of the model ellipsoid, $q$ and $s$. These can be then used to predict projected, observed shapes. We relate $q$ and $s$ to the properties of the embedding halo  throughout the following sections.

\section{Results}
\label{sec:results}

In this section, we focus first on describing the properties of the sample of central galaxies. Their shape misalignments with respect to the host haloes are described in Section \ref{sec:mis}. Section \ref{sec:resid} explores the dependence of the misalignment angle on galaxy properties and environment. The distribution of galaxy shapes is examined in Section \ref{sec:shapes}. We examine the shapes and alignments of satellites in Section \ref{sec:satellites}. Section \ref{sec:corr} presents the orientation correlations of haloes and galaxies, both satellites and centrals, and their cross-correlations. Section \ref{sec:spins} describes the misalignment of galaxy spins with respect to the host halo and presents spin correlations. Finally, Section \ref{sec:nomatch} discusses the contribution of the galaxies that were not matched to haloes in Horizon-DM to intrinsic alignment correlations.

\begin{figure*}
  \centering
  \includegraphics[width=0.99\textwidth]{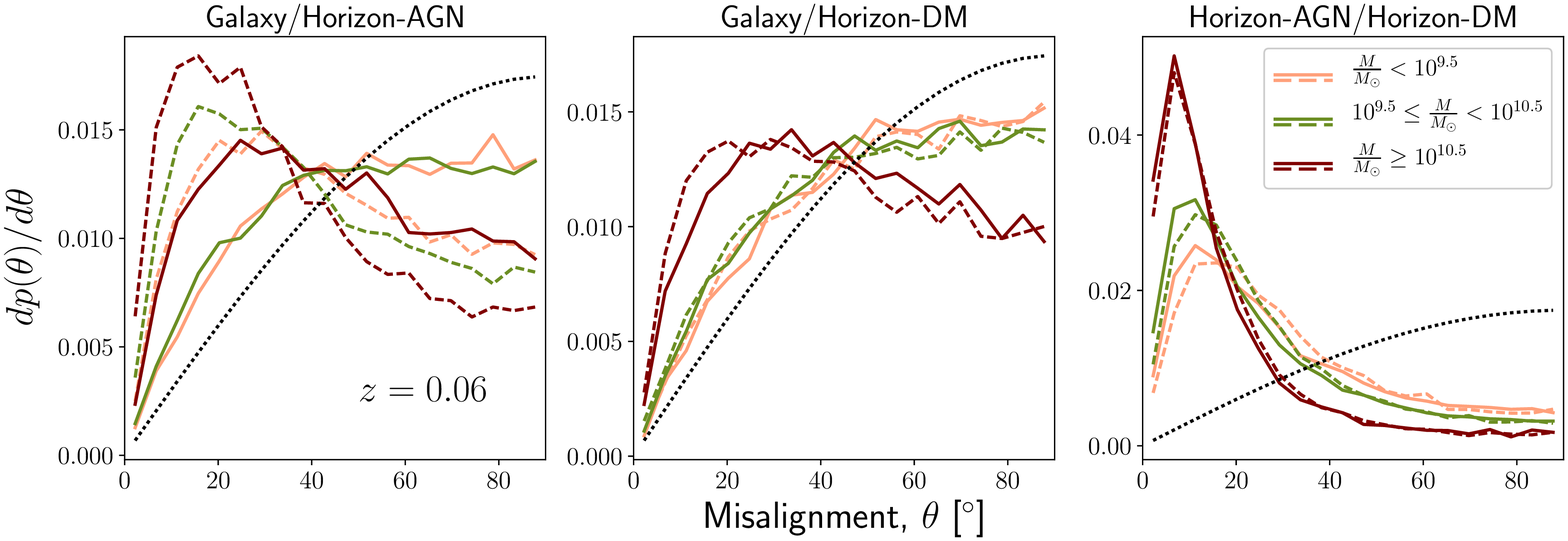}
  \includegraphics[width=0.99\textwidth]{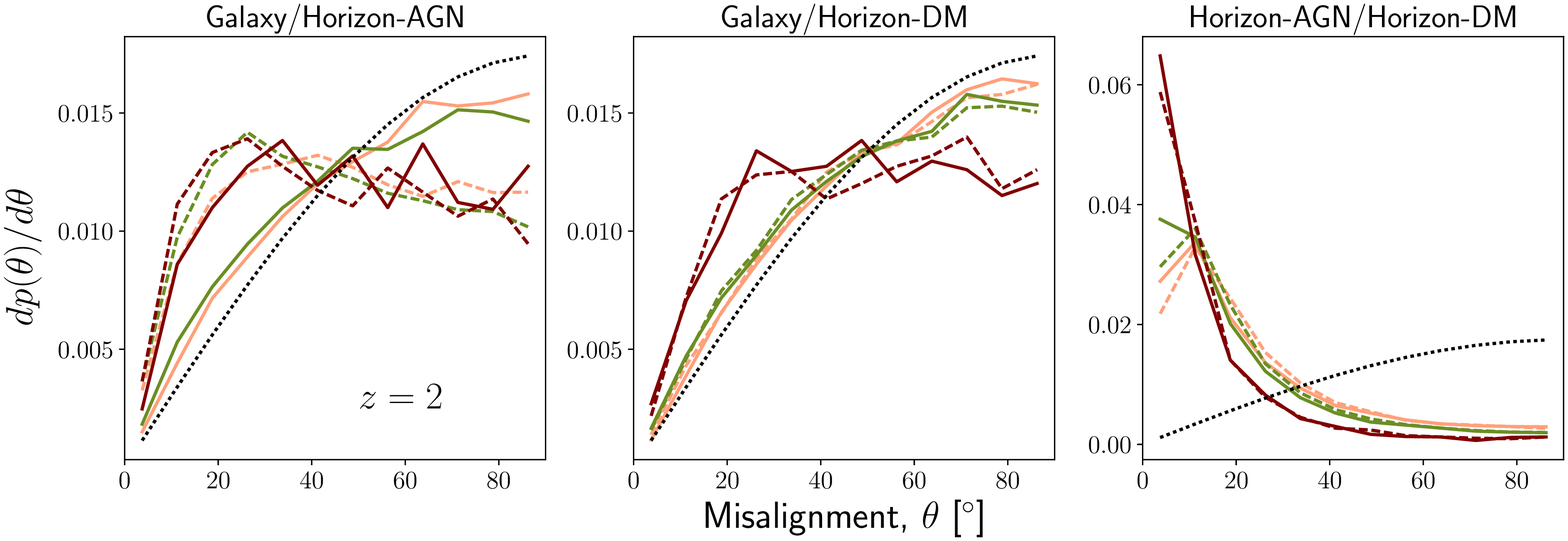}
  \caption{The distribution of misalignment angles between galaxies and haloes at $z=0.06$ (top) and $z=2$ (bottom) in Horizon-AGN (left), of galaxies and haloes in Horizon-DM (middle) and between haloes in both simulation runs (right) for three bins in stellar mass. These correspond to median halo masses of $\langle M_h\rangle = 10^{10.8}$ M$_\odot$, $\langle M_h\rangle = 10^{11.2}$ M$_\odot$ and $\langle M_h\rangle = 10^{12}$ M$_\odot$, from light to dark gray and to black, respectively. Solid curves indicate measurements using the simple inertia tensor ($SIT$) and dashed curves correspond to the reduced inertia tensor ($RIT$). The black dotted line represents the expectation when the angles are distributed at random. Colour figures are available in the online version of the manuscript.}
  \label{fig:align}
\end{figure*}

\subsection{Alignments of galaxies and haloes}
\label{sec:mis}

A key ingredient to halo models of intrinsic alignments is the relative orientation between the galaxy and the halo. In this work, we focus on determining the relative angle between their minor axes. The major axis is not well defined for disc-like structures, where it can become degenerate with the middle axis and thus more prone to noise.

In Figure \ref{fig:align}, we show the distribution of misalignment angles of galaxies with respect to their embedding halo at $z=0.06$ (top) and $z=2$ (bottom) in Horizon-AGN (left panel), and with respect to the matched Horizon-DM halo (middle panel) in the stellar mass bins defined in Section \ref{sec:cats}. The dotted line represents the PDF for random alignments.

The top left panel of Figure \ref{fig:align} shows that galaxy alignments are clearly departing from the random distribution at low redshift and high masses, and particularly for the alignment of the reduced inertia tensor of the galaxy with the reduced inertia tensor of the halo. The outer part of the halo, probed by the simple inertia tensor is less aligned with the galactic simple inertia tensor. This is likely a consequence of the ellipsoidal model of the halo changing orientation towards the outskirts, and the fact that the simple inertia tensor of a halo probes much larger scales than the scale of the galaxy. The mean misalignment between the orientation of the minor axis of a halo using the simple inertia tensor, compared to the reduced inertia tensor, is $19.9^\circ\pm 0.1^\circ$. As some alignment remains even between the galaxy and the halo for the simple inertia tensor, this suggests that the inner and outer region of a halo are not completely decoupled for these shape estimators. All results in Figure \ref{fig:align} are obtained using the sample of galaxies that have been linked to haloes both in Horizon-AGN and Horizon-DM. However, the results are qualitatively the same in the case where we include galaxies that have not been matched to Horizon-DM.

Comparing the top left and bottom left panels of Figure \ref{fig:align}, we find that alignments between galaxies and haloes are less significant at higher redshift. At $z=2$, the alignment  of the reduced inertia tensor shapes is mostly consistent with the random distribution for galaxies of intermediate and low mass. This suggests that some physical process is increasing the alignment between galaxies and haloes as time passes by, and this effect is mass-dependent, with more massive galaxies becoming more aligned.

The right panels of Figure \ref{fig:align} represent the relative misalignment between cross-matched haloes in Horizon-DM and Horizon-AGN. Haloes are generally well-aligned between simulations, with mean misalignment angles of $\{31.5^\circ\pm 0.2^\circ,27.4^\circ\pm 0.1^\circ$,$18.6^\circ\pm 0.2^\circ\}$ from the low mass bin to the high mass bin at $z=0.06$, and $\{24.3^\circ\pm 0.1^\circ,20.5^\circ\pm 0.1^\circ$,$13.6^\circ\pm 0.4^\circ\}$, respectively, at $z=2$. These results are quoted with the standard error of the mean and the simple inertia tensor; very similar results are obtained for the reduced inertia tensor. Thus, baryonic physics are not able to completely de-correlate the dark matter distribution in the haloes between the two simulations. Cross-matched haloes between the twin runs are typically well-aligned with each other, and the misalignment PDF is not very sensitive to the choice of shape estimator, as we find similar misalignment distributions for the minor axes as defined using either the simple or reduced inertia tensors. Nevertheless, as quoted above, the inner and outer regions of individual haloes, as probed by the direction of the minor axis from simple and reduced inertia tensors, can exhibit significant ($\sim 10^\circ-30^\circ$) misalignment. The alignment is a function of mass, with higher mass haloes better preserving their relative alignment. Comparing the top and bottom rows, we see that haloes in the two simulation become more misaligned as time goes by, due to the cumulative effect of baryons on their orientations. 

\begin{figure}
  \centering
  \includegraphics[width=0.47\textwidth]{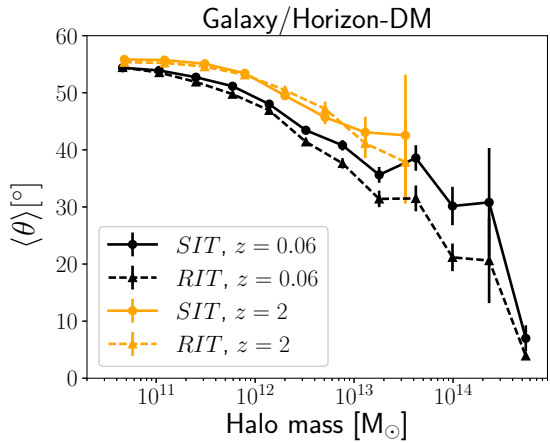}
  \caption{The average misalignment angle between galaxies and haloes in Horizon-DM as a function of the mass of the halo at $z=0.06$ (black) and $z=2$ (orange in online version, light gray in printed version). The dashed curves correspond to the misalignment angle between reduced inertia tensors ($RIT$). The solid curves, to the misalignment angle between simple inertia tensors ($SIT$). The average misalignment is a decreasing function of mass and redshift. The error bars are the standard error about the mean of each bin. The last bin only contains a few galaxies which implies the error may be underestimated.}
  \label{fig:avgmis}
\end{figure}
\begin{figure*}
  \centering
  \includegraphics[width=0.99\textwidth]{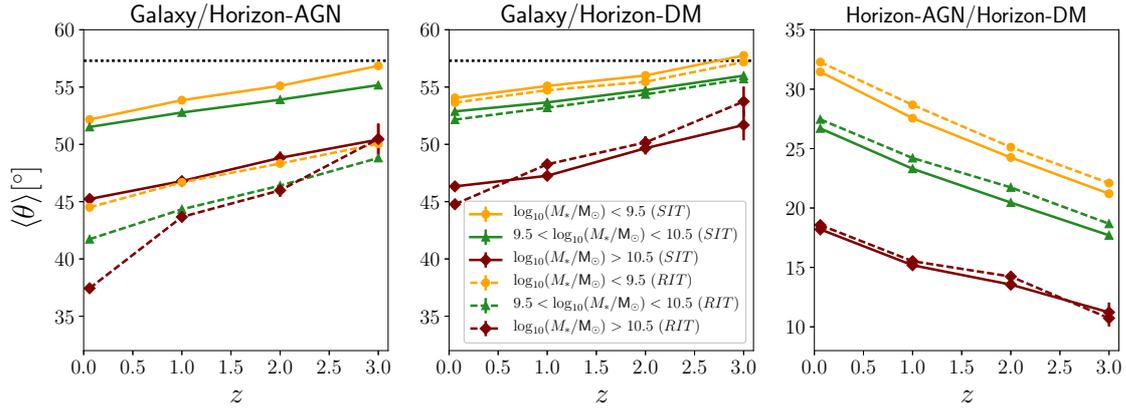}
  \caption{The redshift evolution of the mean misalignment angle between galaxies and haloes in Horizon-AGN (left panel), between galaxies and haloes in Horizon-DM (middle panel) and between haloes in both runs (right panel). The solid curves correspond to the simple inertia tensor ($SIT$); the dashed curves, to the reduced inertia tensor ($RIT$). The results are presented for three different ranges of stellar mass, corresponding to different symbols as shown in the legend.}
  \label{fig:misevol}
\end{figure*}
\begin{figure}
  \centering
  \includegraphics[width=0.49\textwidth]{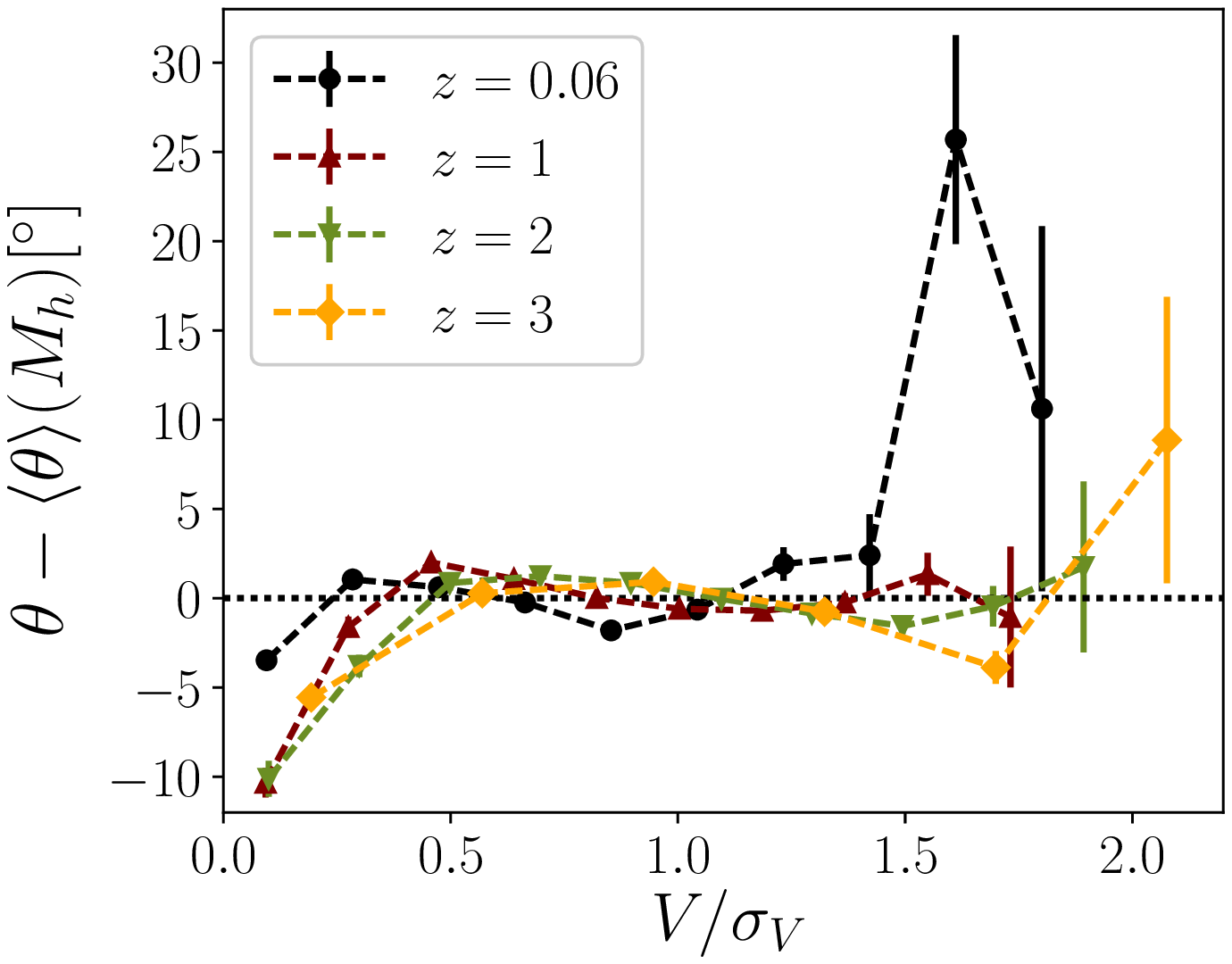}
  \caption{The misalignment angle residuals computed from the reduced inertia tensor shapes as a function of $V/\sigma_V$ of the galaxy at different redshifts in the range $0<z<3$ (lighter gray indicates increasing redshift). There are significant residuals for disc galaxies at low redshift; and for ellipticals at high redshift. A dependence of galaxy-halo alignment on halo mass might not be sufficient for prescribing galaxy orientation.}
  \label{fig:resid}
\end{figure}

We have also measured the angle between the minor axis of each galaxy and its matched Horizon-DM halo. This is shown in the middle panels of Figure \ref{fig:align}, where the misalignment is larger than for the left panels. This is in part due to a small halo-halo misalignment between the two runs (right panels). The middle panels can be seen as an approximate convolution of the left and right panels. This PDF is of interest if one were to populate DMO simulations using a halo model of galaxy alignment.

Note that to describe the relation between the three minor axes shown in Figure \ref{fig:cartoon}, a piece of information is yet missing. Apart from the distribution of misalignment angles between galaxy and halo, and between haloes, we should also characterise the distribution function for the angle $\phi$, the angle between the minor axis of the galaxy and the plane determined by the minor axes of the matched haloes. While this information is not needed to populate a DMO simulation with aligned galaxies (in this case, only the PDF of $\theta$ is required), we find that the mean $\langle \phi \rangle$ is typically $\sim 65.5^\circ \pm 0.1^\circ$ (standard error of the mean quoted), well above the random expectation of $57^\circ$ for galaxies at all masses. In other words, the three minor axes considered are preferentially aligned with each other.

Figure \ref{fig:avgmis} shows\footnote{Colour versions of the figures are availables in the online version of this manuscript.} the average misalignment between central galaxies and haloes in Horizon-DM as a function of halo mass at two redshifts and for the two different shape estimators. The mean misalignment angle is a function of mass, with more massive haloes hosting more aligned galaxies. At $z=2$ (orange/light gray), the alignment is lower than at $z=0.06$ (black), but shows similar mass dependence. In Figure \ref{fig:misevol}, we show the redshift evolution of the mean misalignment angle for the three different stellar mass ranges. The mean misalignment angle between galaxies and haloes increases towards high redshift, approaching the random expectation (dotted line). Haloes, on the other hand, increase their mean misalignment towards low redshift, likely as the consequence of the cumulative effect of baryonic feedback processes.

We also find that there is always more alignment between reduced inertia tensors of galaxy and halo in the Horizon-AGN simulation than simple inertia tensors. This difference is mainly due to the larger physical extent of haloes. Because galaxies occupy a small sub-volume within the halo they inhabit, the alignment of the galaxy shape with the halo shape is stronger if the reduced inertia tensor of the halo is considered. In addition, there is observational motivation to use the reduced inertia tensor of a galaxy as a proxy for the observed galaxy shape, since shape measurement methods optimised for weak lensing often up-weight the more luminous inner region of a galaxy. In practice, though, Figure \ref{fig:align} shows that the difference between simple and reduced inertia tensor for determining relative orientations is significantly reduced when comparing galaxy and Horizon-DM halo orientations. Notice that this is not the case when determining galaxy ellipticities in Section \ref{sec:shapes}.

\begin{figure*}
  \centering
  \includegraphics[width=0.49\textwidth]{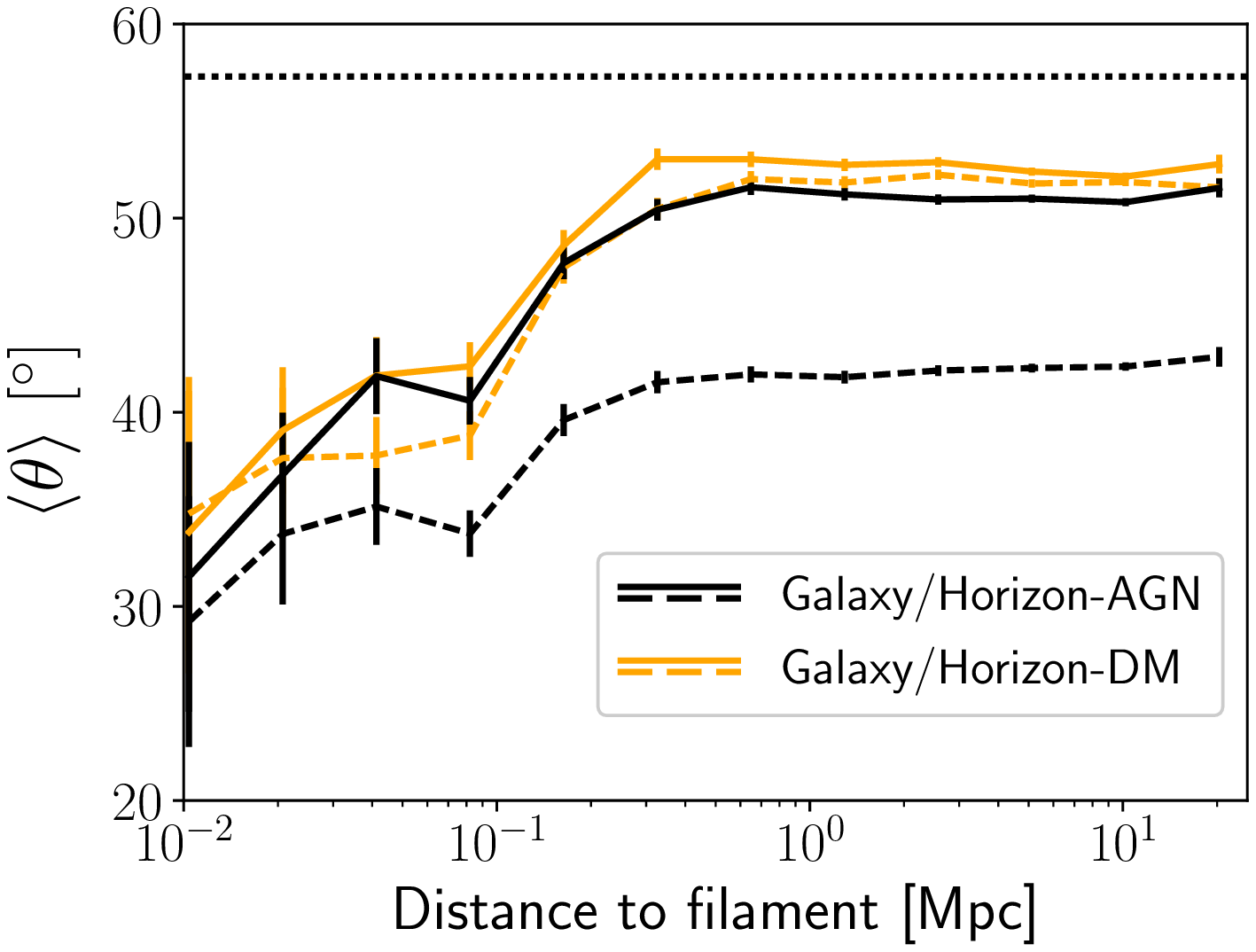}
  \includegraphics[width=0.49\textwidth]{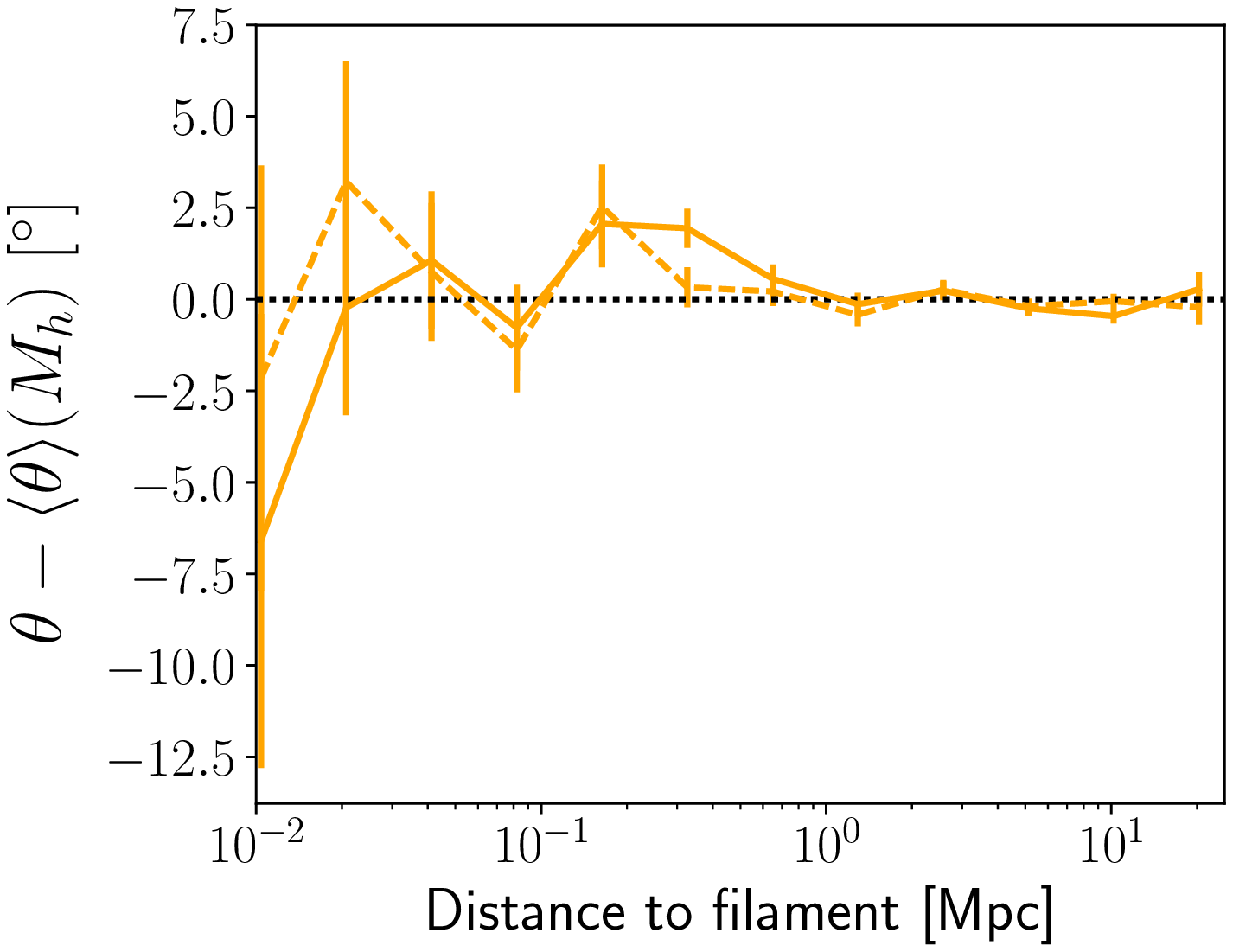}
  \caption{The left panel presents the misalignment angle between galaxy and halo minor axes as a function of distance to the closest filament at $z=0.06$. Dashed curves represent results for the reduced inertia tensor; solid curves, for the simple inertia tensor. The mean misalignment with respect to haloes in Horizon-AGN is shown in black; and for matched haloes in Horizon-DM, in orange (online version) or light gray (printed version). At small distance from the filament, haloes and galaxies are better aligned with each other. In the right panel, we see that the residuals from the mean $\langle \theta \rangle (M_h)$ relations from Horizon-DM bear no relation to the distance to the nearest filament. The dependence on halo mass fully captures the dependence seen in the left panel.}
  \label{fig:filament}
\end{figure*}

\subsection{Dependence of alignment on galaxy properties and environment}
\label{sec:resid}

While we have shown that the misalignment angle is dependent on halo mass, if we want to build a prescription for populating DMO simulations with galaxy alignments, we should check whether the mean misalignment angle correlates with other variables. Intrinsic alignment models in the literature distinguish different mechanisms for alignment depending on whether a galaxy is pressure-supported or whether it has a significant angular momentum \citep{Catelan01}. It is also known that mergers have an impact on determining morphological transitions and spin re-orientations \citep{Codis12,Dubois14,Cen14,welker2014}. Hence, there could be additional dependence of the mean misalignment angle on $V/\sigma_V$ or on environment. We look for those potential dependencies in this section. We show results only for the case of the reduced inertia tensor, though we obtain similar results for the simple inertia tensor.

We compute residuals with respect to the mean misalignment as a function of mass, which was shown in Figure \ref{fig:avgmis}. We test whether these residuals, $\Delta\theta=\theta-\langle \theta \rangle (M_h)$, are correlated with the dynamical properties of the galaxies, as encoded by $V/\sigma_V$. The results are shown in Figure \ref{fig:resid}. At $z=0.06$, our results suggest the presence of additional dependence of the misalignment angle at fixed mass for disc-like galaxies.  The alignment signal depends on the formation history, and hence the dynamical properties of a galaxy at fixed mass. At higher redshift, on the other hand, we find that the $\langle \theta \rangle (M_h)$ relation fails to predict the alignment of ellipticals.

Similarly, it is possible that the galaxy-halo misalignment angle is related to the cosmic web environment. Environmental dependence of the alignment signal has been proposed in several works \citep{Dubois14,welker2014,codis15b,Welker15,Laigle15}. We explore whether the relative orientation of galaxies and haloes depend on their specific relation to the cosmic web, characterised by the distance to the nearest filament and the direction of the closest filament. We find that the misalignment angle between a galaxy and its host halo depends on the distance to the nearest filament, as shown in the left panel of Figure \ref{fig:filament}. Haloes closer to a filament are better aligned with cross-matched galaxies. The right panel of Figure \ref{fig:filament} shows that the environment has no other specific impact on the relative alignment of a galaxy and a halo than the one seen through the halo mass. Although not shown, we also find that the direction of the misalignment is unrelated to the direction of the filament.

\begin{figure*}
  \centering
  \includegraphics[width=0.49\textwidth]{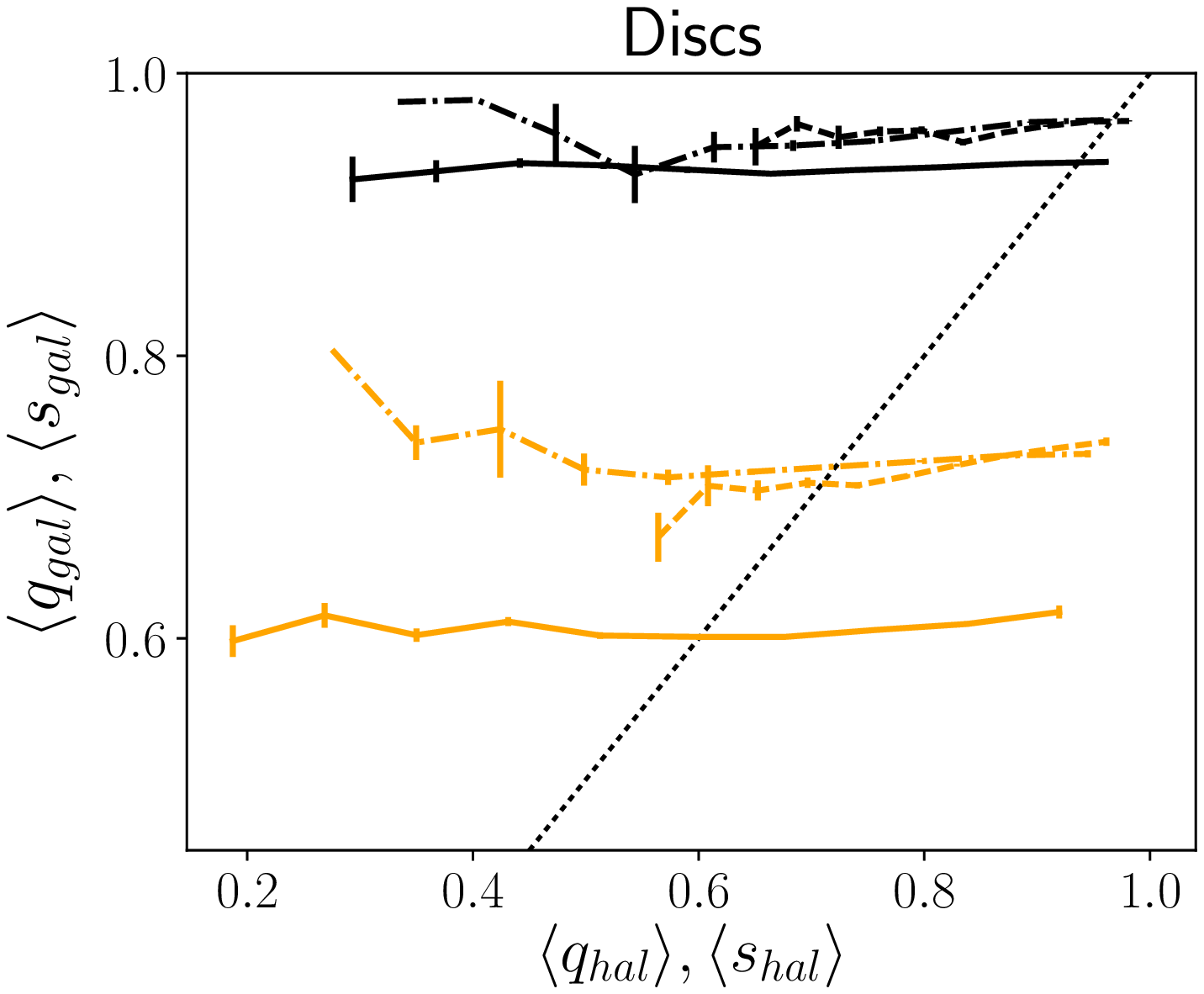}
  \includegraphics[width=0.49\textwidth]{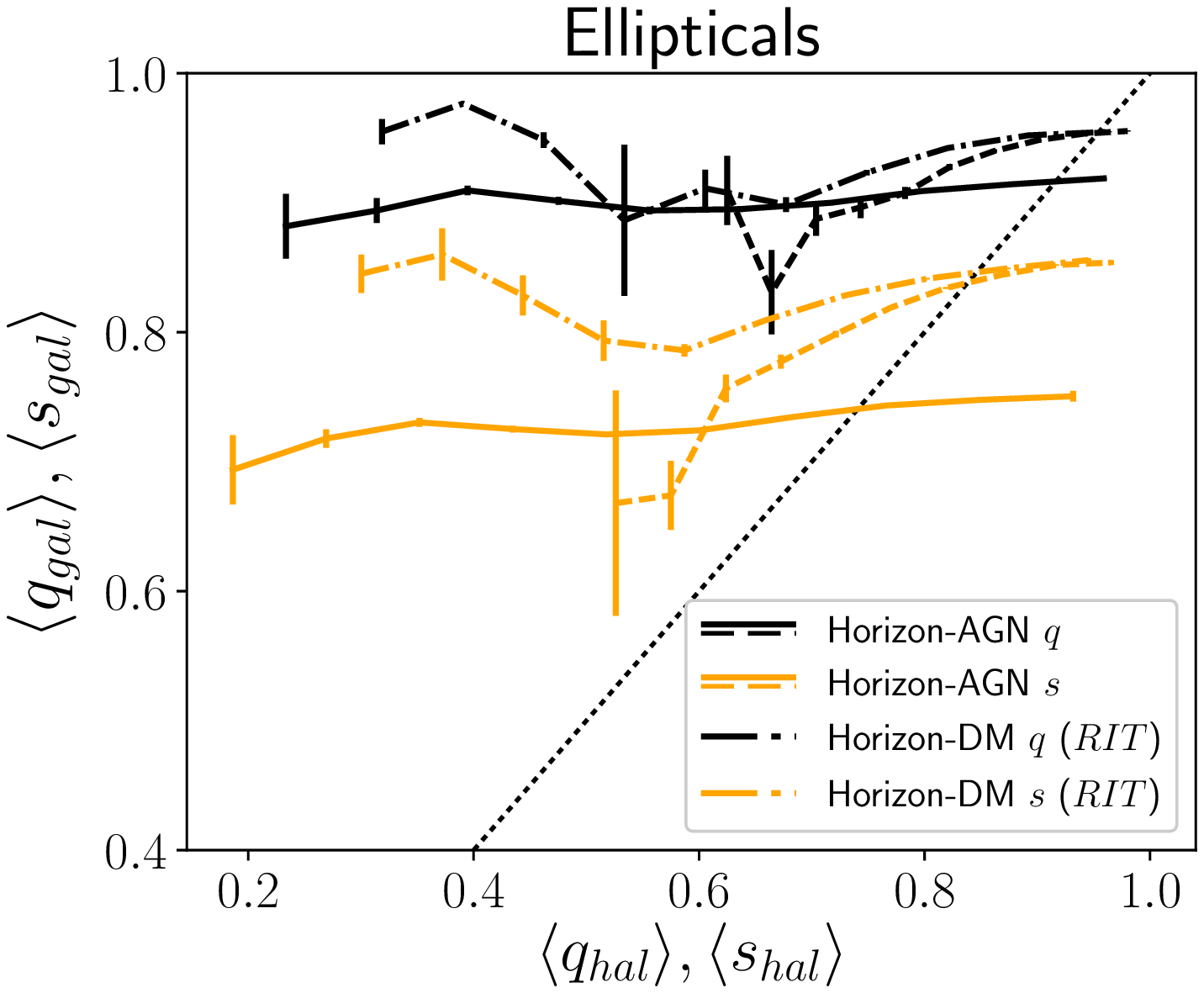}
  \caption{The axis ratios of discs (left panel) and ellipticals (right panel) at $z=0.06$ shown in black ($q$) and orange/light gray ($s$). These are compared to the axis ratios of haloes in Horizon-AGN (solid for simple inertia tensor, $SIT$, and dashed for reduced inertia tensor, $RIT$) and Horizon-DM (dot-dashed). The dotted curves indicate the identity. Disc and elliptical axis ratios cannot be predicted from the shape of the matched Horizon-DM halo.}
  \label{fig:axescorr}
\end{figure*}
\begin{figure*}
  \centering
  \includegraphics[width=0.42\textwidth]{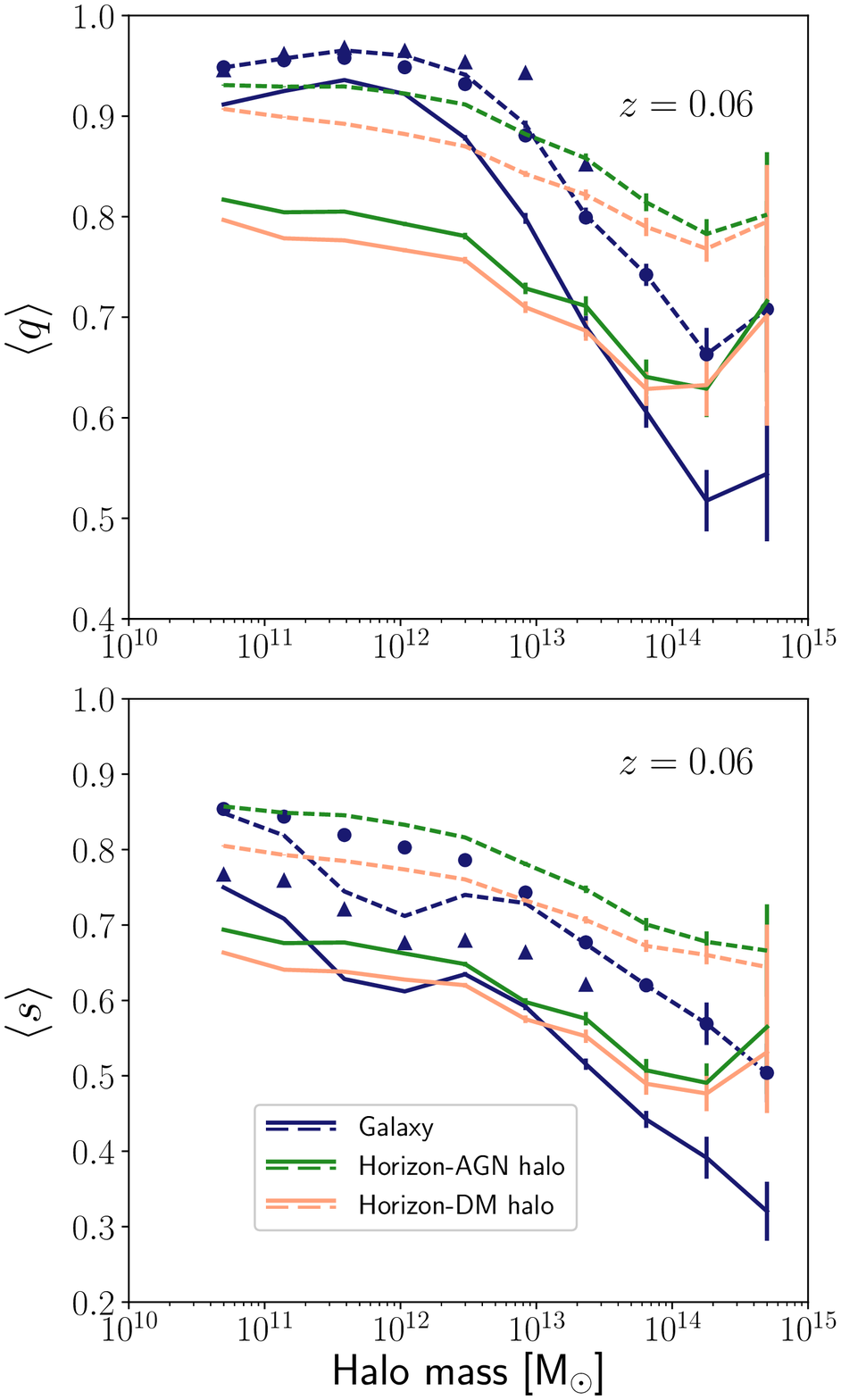}
  \includegraphics[width=0.42\textwidth]{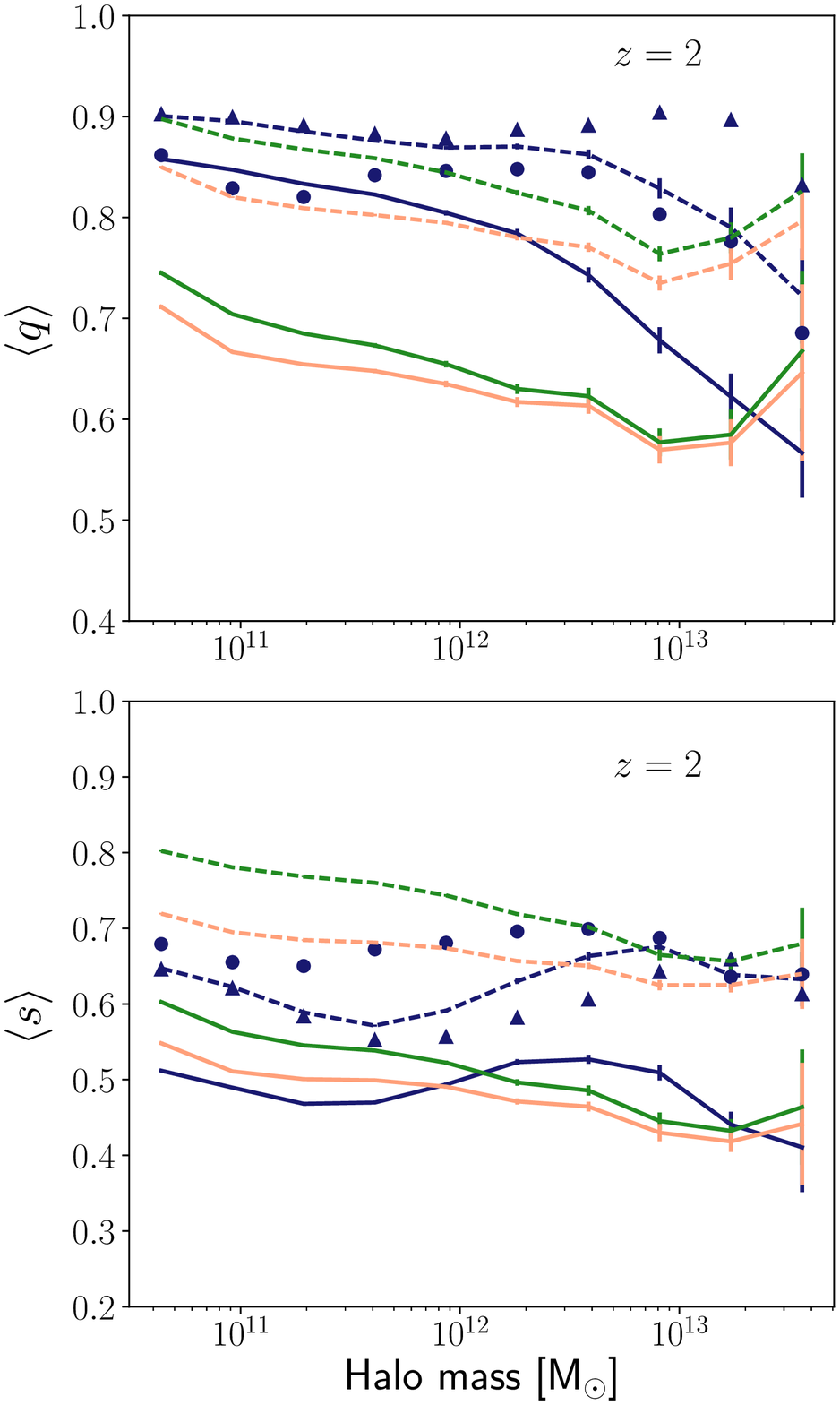}
  \caption{The average axis ratios of galaxies and haloes as a function of the halo mass at $z=0.06$ (left) and $z=2$ (right). The top panel shows the average $q$, the ratio of the middle to the major axes. The bottom panel shows the average $s$, the ratio of the minor to major axis. The solid curves represent the measurements from the simple inertia tensor ($SIT$); the dashed curves, from the reduced inertia tensor ($RIT$). Blue/black corresponds to central galaxies, green/dark gray corresponds to their matching dark matter haloes in Horizon-AGN and orange/light gray to those in Horizon-DM. The blue line in the $RIT$ implementation is also split into discs (triangles) and ellipticals (filled circles). The error bars are the standard error of the mean in each bin. Overall, there is a tendency for both galaxies and haloes to become more flattened at high masses. }
  \label{fig:axesratio}
\end{figure*}

\subsection{Shapes of galaxies and haloes}
\label{sec:shapes}

Can the shape of a galaxy be related to that of the host halo? We directly test this assumption by comparing galaxy and halo axis ratios in Figure \ref{fig:axescorr}, where we divide the galaxy population into discs (left panel) and ellipticals (right panel) at $z=0.06$. This figure shows that there is no correlation between the axis ratios of discs and those of the embedding haloes in Horizon-AGN for either reduced or simple inertia tensor measurements. This result holds as we consider the axis ratios of the cross-matched Horizon-DM halo. In the case of ellipticals, the right panel of Figure \ref{fig:axescorr} shows that their axis ratios cannot be related to those of the halo in the case of the simple inertia tensor. There is a correlation between axis ratios of an elliptical and the Horizon-AGN halo when the reduced inertia tensor is considered, although it is not possible to directly assign a galaxy the shape of the embedding halo (dotted line). This correlation is lost in the comparison with the matched Horizon-DM halo, suggesting that ultimately it is impossible to use the inertia tensor of a halo in an DMO simulation to predict that of any galaxy.   

What are the typical halo and galaxy shapes and can they be predicted from the halo mass? In Figure \ref{fig:axesratio}, we show the average axis ratios, $q$ and $s$ (defined in Section \ref{sec:shapedef}), for galaxies and haloes as a function of halo mass at $z=0.06$ (left panel) and $z=2$ (right panel). At $z=0.06$, we find overall that lower mass haloes, and galaxies inhabiting those haloes, are rounder than higher mass counterparts. This trend is visible for both simple and reduced inertia tensors. However, reduced inertia tensor ratios are overall closer to $1$, in part due to the symmetry imposed by the spherical weighting of this measurement. This figure once again confirms that there is little correlation between the axis ratio of a galaxy and that of its embedding halo. Thus, the dark matter halo shape cannot be used to predict the galaxy shape. Similar conclusions were reached by \citet{Suto16} in an analysis of galaxy clusters in Horizon-AGN. Unfortunately, the curves in Figure \ref{fig:axesratio} cannot be used to predict the shape of a galaxy given the halo mass either. This is due to significant dependence of the axis ratios on galaxy morphology. Splitting the population in discs and ellipticals, we find significant departures from the mean trend in Figure \ref{fig:axesratio}. 

Both haloes and galaxies become more elliptical at high redshift. This is evidenced by the lower values of $\langle q \rangle$ and $\langle s \rangle$ in the right panel of Figure \ref{fig:axesratio}. Consistent with this phenomenon, the difference between the average axis ratios estimated by the simple inertia tensor and the reduced inertia tensor increases towards high redshift. We also find that haloes in Horizon-AGN are more spherical than those in Horizon-DM; thus the effect of baryons is to produce rounder haloes. 

\subsection{Satellite shapes and alignments}
\label{sec:satellites}

In Section \ref{sec:matching}, we described the selection of the satellite galaxy sample. Considering that DMO simulations can only have access to information encoded in the dark matter field, we define satellites to be those galaxies in Horizon-AGN that have been cross-matched to a subhalo in Horizon-DM. The fraction of satellites in our total sample goes from $28\%$ at $z=0.06$ to $18\%$ at $z=3$, where we have about $6000$ satellites. These satellite samples are large enough to study some of the statistical properties of satellite shapes and alignments. 

\begin{figure*}
  \centering
  \includegraphics[width=0.49\textwidth]{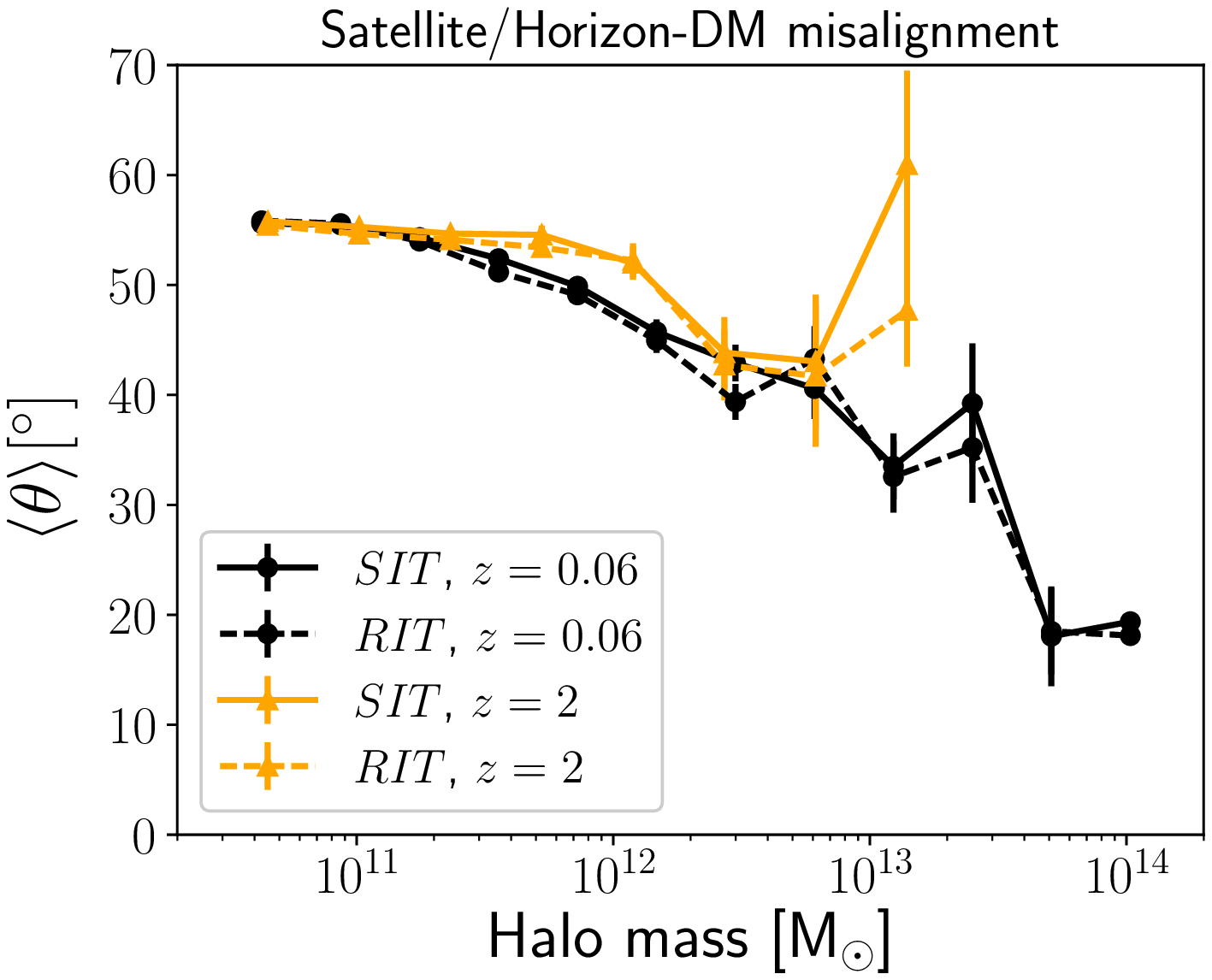}
  \includegraphics[width=0.49\textwidth]{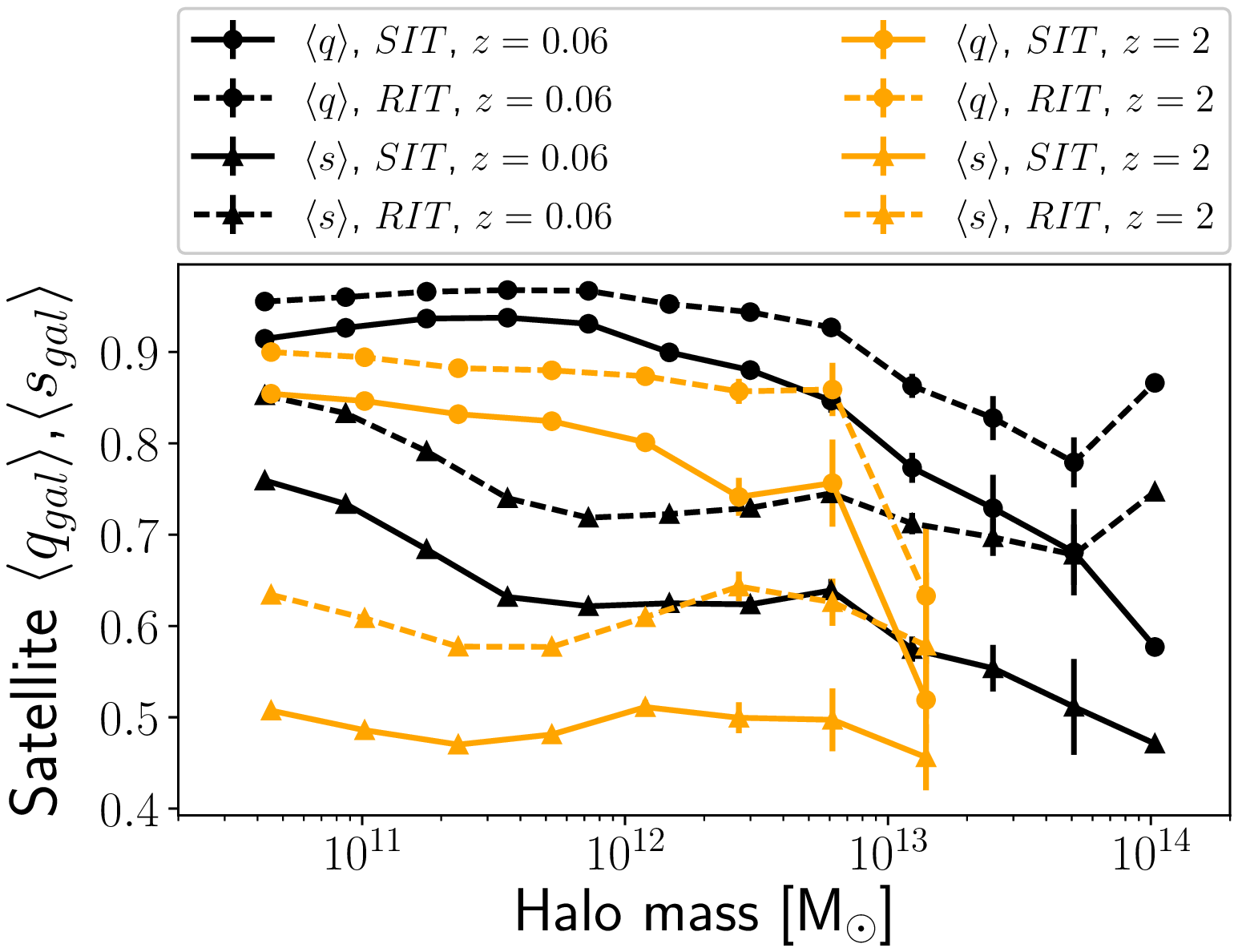}
  \includegraphics[width=0.49\textwidth]{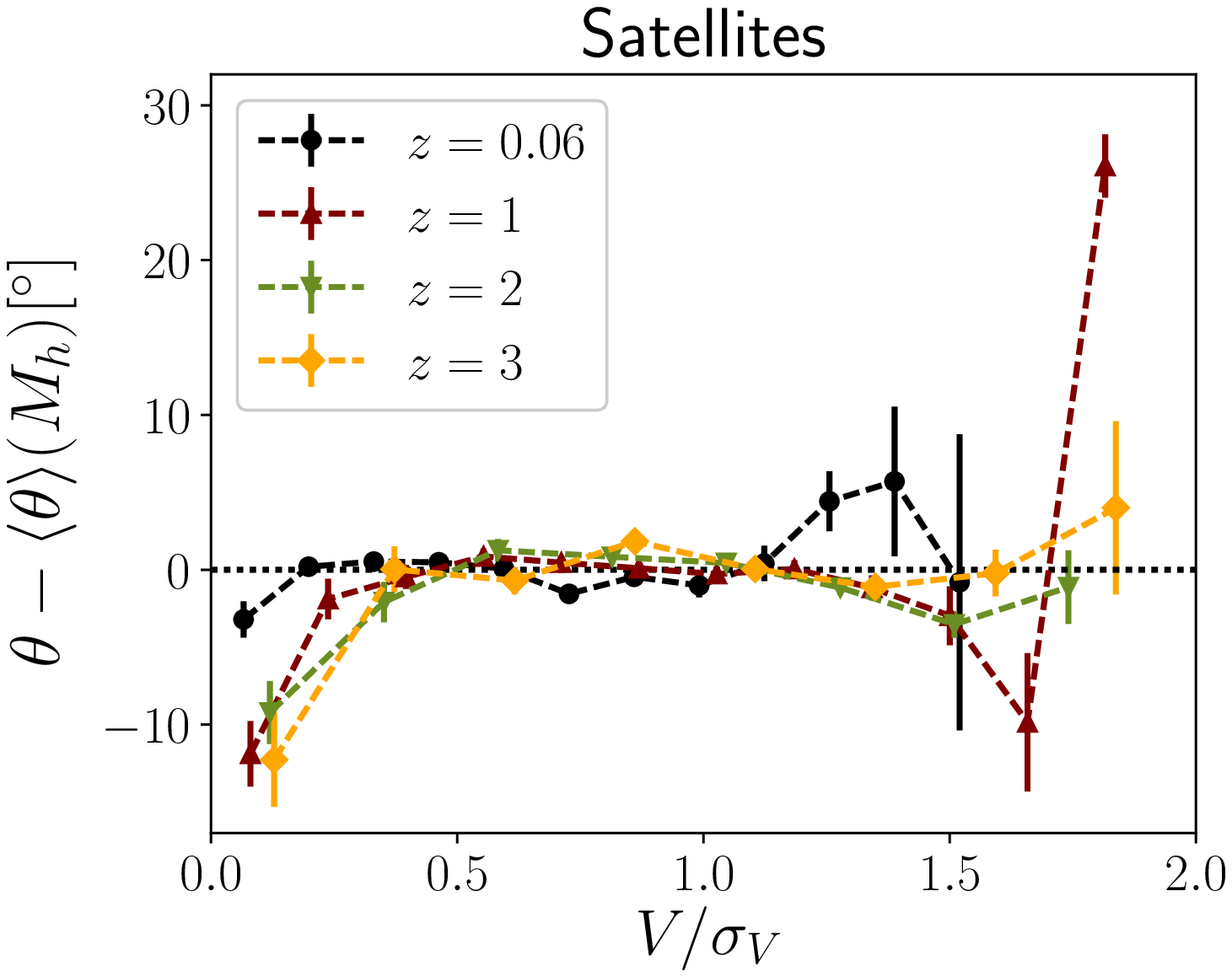}
  \includegraphics[width=0.49\textwidth]{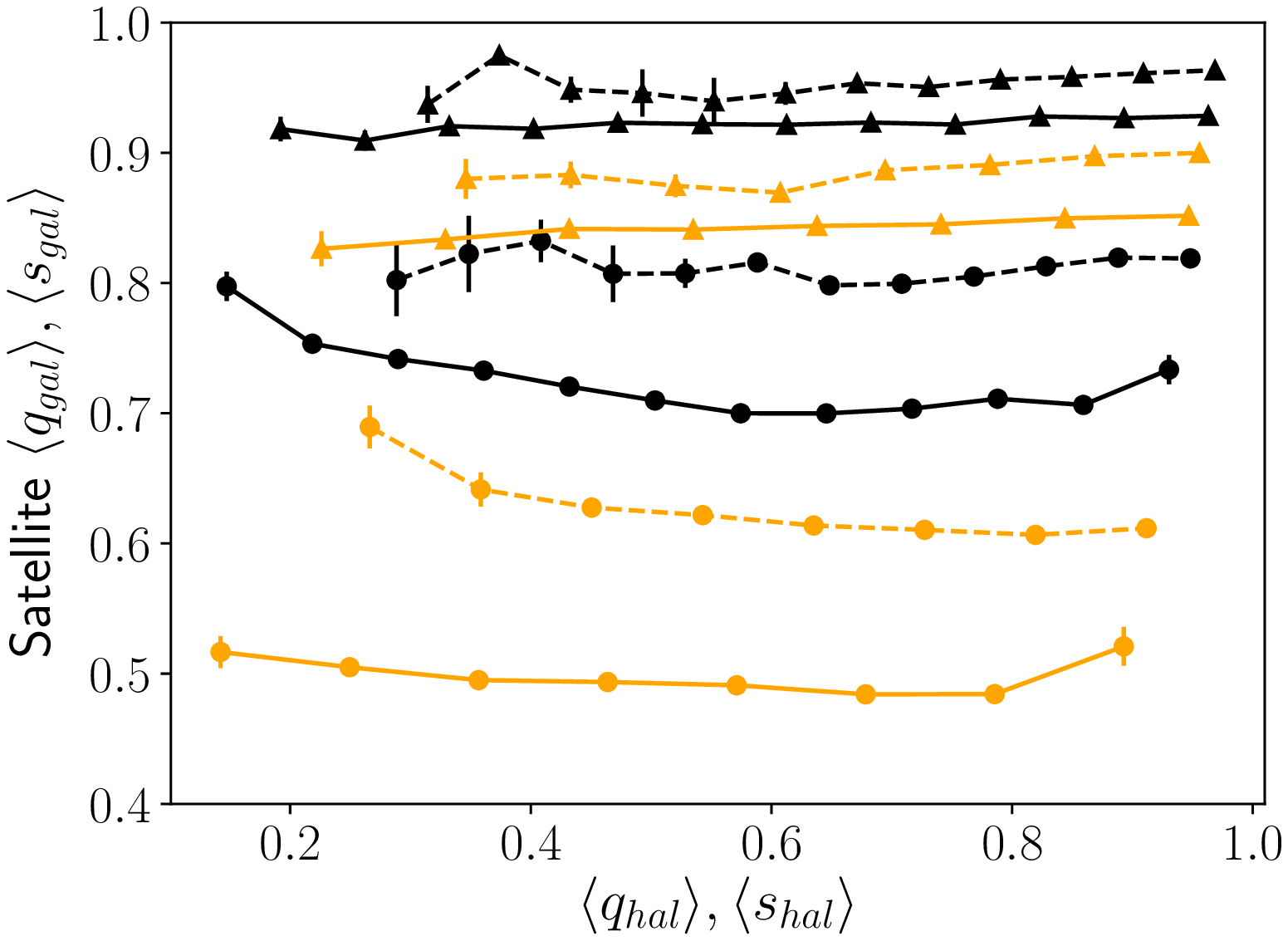}
  \caption{This figure summarizes the shapes and orientations of satellite galaxies in Horizon-AGN with respect to the matched subhaloes in Horizon-DM. The top left panel represents the mean misalignment angle between the minor axis of a satellite and the matched Horizon-DM subhalo for simple (solid) and reduced (dashed) inertia tensor at $z=\{0.06,1,2,3\}$ (black to lightest gray). The top right panel shows the mean axis ratios of satellite galaxies as a function of halo mass at those two redshifts, with a similar format adopted. In this panel, $\langle q\rangle$ is represented by the filled circles; $\langle s\rangle$, by filled triangles. The bottom right panel shows the relation between the satellite axis ratios and the host halo axis ratios (with the same colour and symbol scheme as the top left panel). Finally, the bottom left panel shows the departure from the predicted $\langle \theta \rangle (M_h)$ angle as a function of $V/\sigma_V$ at different redshifts.} 
  \label{fig:satell}
\end{figure*}

Figure \ref{fig:satell} summarizes our results for satellite-halo alignments. The top left panel shows the average misalignment angle between a galaxy and the matching Horizon-DM halo at two different redshifts and for the two shape measurement methods. We find similar trends to those of central galaxies, namely that the mean misalignment angle decreases with mass, and increases with redshift. Results are very similar for simple and reduced inertia tensor.

The top right panel of Figure \ref{fig:satell} shows the average axis ratios for satellite galaxies as a function of Horizon-DM halo mass. Galaxies become rounder at lower redshift, and we find significant difference between the simple and reduced inertia tensor. As expected, reduced inertia tensor shapes are rounder than simple inertia tensor shapes. At low redshift, there is a significant trend with halo mass, similar to the trend observed for centrals.

The bottom left panel shows the departures from $\langle \theta \rangle (M_h)$ for satellite galaxies as a function of $V/\sigma_V$. At $z=2$, the misalignment angle of ellipticals can be offset from the predicted $\langle \theta \rangle (M_h)$ by several degrees, similarly to Figure \ref{fig:resid}. Finally, the bottom right panel shows that satellite axis ratios are uncorrelated with the axis ratios of their host haloes, as was the case for centrals in the previous subsection.

\subsection{Halo and galaxy correlations}
\label{sec:corr}

\begin{figure*}
  \centering
  \includegraphics[width=0.95\textwidth]{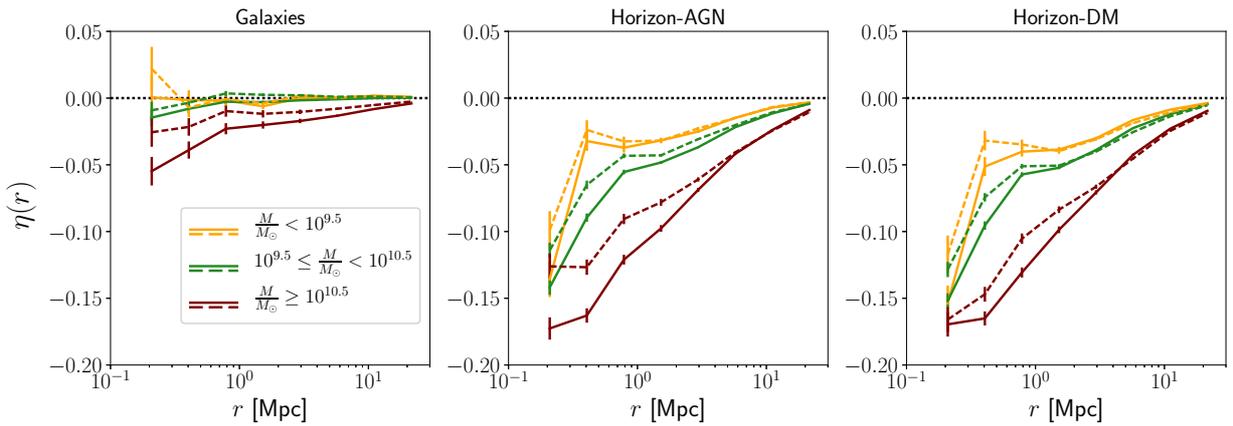}
  \caption{The position-orientation correlation function, $\eta(r)$  (defined in Equation \ref{eq:eta}), for galaxies (centrals and satellites) in different mass bins (left), their matching haloes in Horizon-AGN (middle) and in Horizon-DM (right). The error bars correspond to the variance in each bin. Red/black represents galaxies with $M>10^{10.5}{\rm M}_\odot$, green/dark gray corresponds to galaxies with $10^{9.5}{\rm M}_\odot<M<10^{10.5}{\rm M}_\odot$, and orange/light gray, to galaxies with $M<10^{9.5}{\rm M}_\odot$. A negative signal corresponds to the minor axes being perpendicular to the separation vector. Haloes show stronger alignments than galaxies, particularly in Horizon-DM. A clear mass dependence is seen across all panels.}
  \label{fig:cos2}
\end{figure*}

In the previous section, we presented misalignment angle distributions for galaxies with respect to Horizon-DM matched haloes, and we showed that the misalignment angle is a function of halo mass. If haloes are themselves aligned with each other as a consequence of tidal interactions across the large-scale structure, galaxies ``painted'' onto them will be aligned as well. In this section, we obtain the $\eta$ statistic presented in Eq. (\ref{eq:eta}) for both galaxies and haloes. Results are shown in Figure \ref{fig:cos2} for the combined sample of centrals and satellites; the difference between these two populations will be described below. The left panel shows how the alignment between pairs of galaxies changes with separation. In this panel,  only high mass galaxies show an alignment trend with the large-scale structure, with a preferential perpendicular alignment between the minor axis and the separation vector between galaxies. 

The middle and right panels of Fig. \ref{fig:cos2} present $\eta(r)$ calculated for neighbouring haloes in Horizon-AGN and Horizon-DM, respectively. Haloes have been divided into three bins according to the mass of the matching galaxy. All haloes show significant alignment and there is a clear mass dependence. The alignment signal is present for the least massive haloes, although the galaxies they host have no alignment within our error bars. The relative misalignment between galaxies and haloes results in a smaller alignment strength for galaxies with the large-scale structure. 

The right panel of Figure \ref{fig:align} showed that baryons cause a mass-dependent relative misalignment of matched haloes between the Horizon-DM and Horizon-AGN simulation. This translates into a fractional change of $\eta(r)$ of $10-30\%$ in the high mass range considered, and lower for less massive haloes. The change is more significant in the case of the reduced inertia tensor, suggesting that baryons modify the inner region of the halo more strongly. Note, though, that we are eventually interested in the prediction of galaxy alignments, so as long as such a recipe can be found in connection to the DMO simulation, we do not need to take into account the alignment of haloes in Horizon-AGN. 

\begin{figure*}
  \centering
  \includegraphics[width=0.95\textwidth]{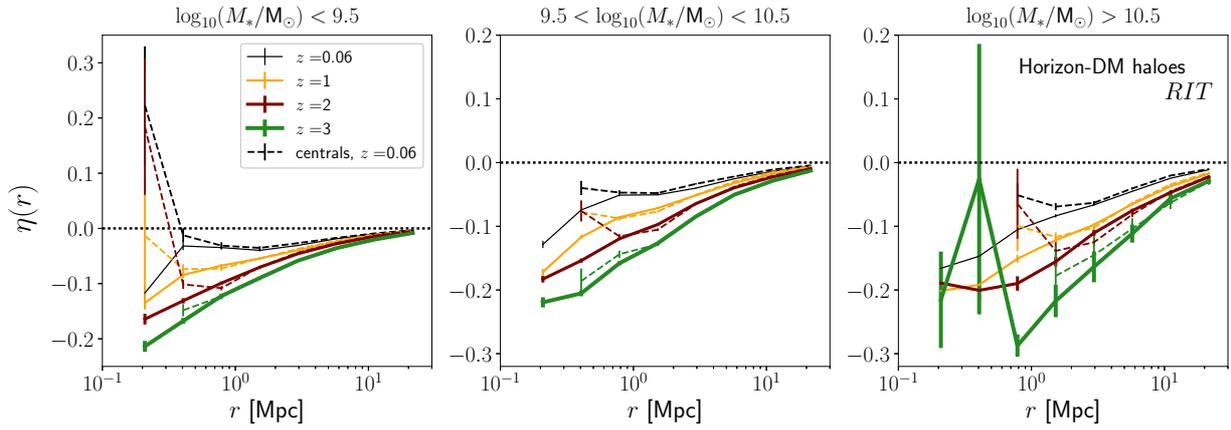}
  \caption{The redshift evolution of $\eta(r)$ (Equation \ref{eq:eta}) for haloes in Horizon-DM inhabited by low mass galaxies (left panel), intermediate mass galaxies (middle panel) and high mass galaxies (right panel). The solid curves correspond to the full sample of centrals and satellites, while the dashed curves highlight the contributions of the centrals alone. We only show results for the reduced inertia tensor with redshift increasing with thickness. The large-scale signal converges to the central-central correlation, while satellites contribute mostly in the one-halo regime. $\eta(r)$ clearly decreases with decreasing redshift. }
  \label{fig:hzdmevol}
\end{figure*}

Figure \ref{fig:hzdmevol} shows the redshift evolution of halo alignments in Horizon-DM. The trends are similar, with lower amplitude, for Horizon-AGN. The three panels in that figure correspond to haloes which host galaxies of different masses. In all cases, $\eta$ is below zero, suggesting that the minor axes of different haloes tend to point perpendicularly with respect to the separation vector. This correlation decreases towards large scales and towards low redshift. Horizon-DM haloes are more strongly aligned at high redshift, and tend to decrease their correlation with time. The convergence of solid and dashed curves at large scales indicates that this regime is dominated by the central-central correlation (dashed), while there is an additional contribution from satellites at small scales (satellite-satellite correlations and central-satellite cross-correlations). Overall, haloes are always more strongly aligned towards each other than galaxies are to other galaxies. While they tend to lose this alignment towards low redshift, the galaxies compete with this effect by aligning their minor axis with that of the halo (left and middle panel of Figure \ref{fig:avgmis}). This has relevant implications for halo models of alignments, which we discuss in the next section. 

\begin{figure*}
  \centering
  \includegraphics[width=0.9\textwidth]{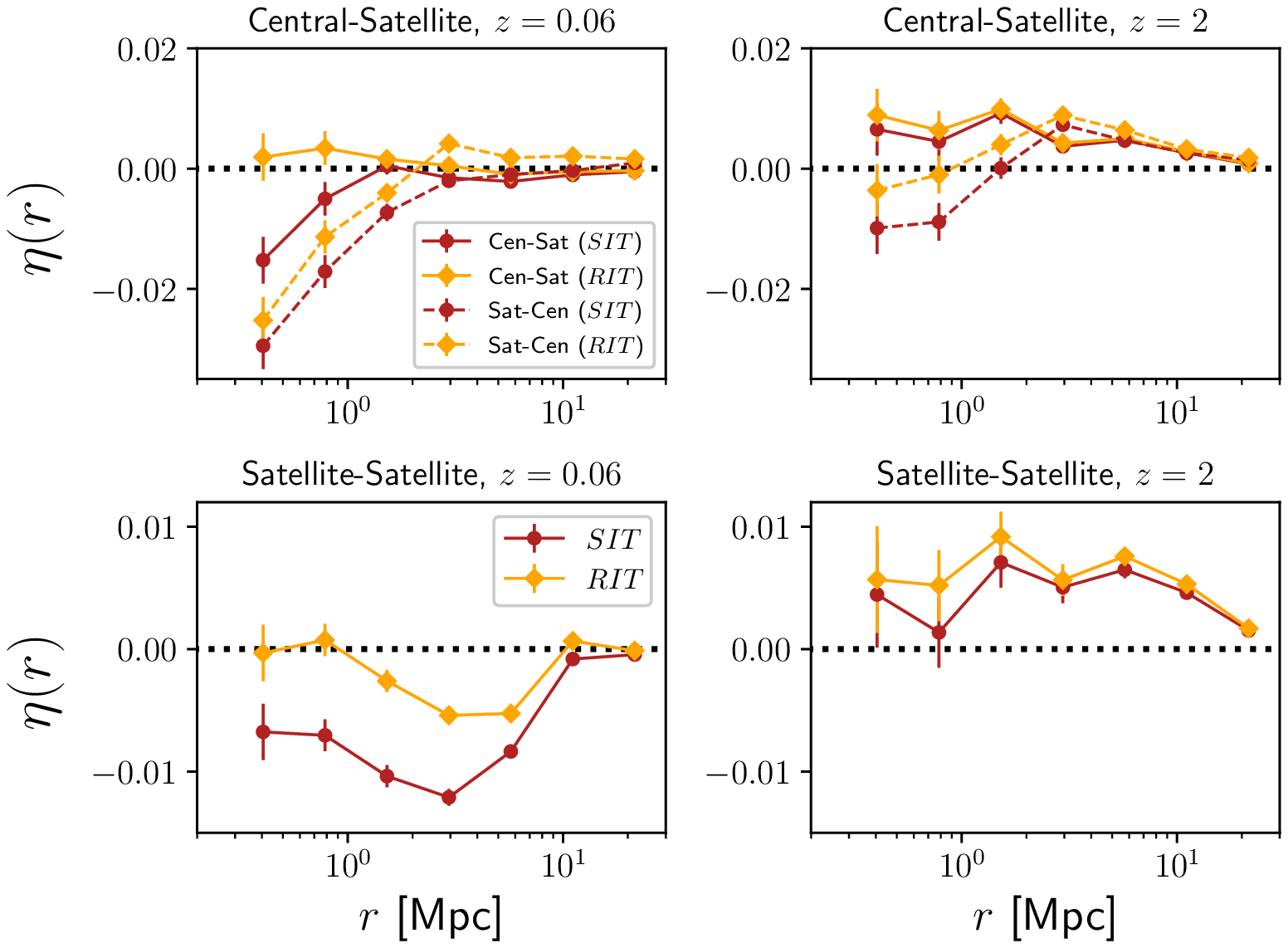}
  \caption{The redshift evolution of the position-orientation correlation, $\eta(r)$ (Equation \ref{eq:eta}), for satellite galaxy alignments. The top panels show the contribution from the relative orientation of the minor axis of the central galaxy and the separation vector to satellite galaxies (dashed), and from the orientation of satellites relative to the location of centrals (solid) at $z=0.06$ (left) and $z=2$ (right). The bottom panels show the satellite-satellite alignment at the same redshifts. Different colours indicate results corresponding to the simple (red/black) and reduced (orange/gray) inertia tensors.}
  \label{fig:satcos2}
\end{figure*}

Correlations between satellites, and between satellites and centrals, contribute to the intrinsic alignment signal at small scales. We show all these contributions at $z=0.06$ and $z=2$ in Figure \ref{fig:satcos2}. The top panels show the $\eta(r)$ statistic for the minor axis of satellites around centrals (solid) and the minor axis of centrals with respect to the location of satellites (dashed). The latter, which represents an anisotropic clustering of satellites in the direction perpendicular to the minor axis of the central galaxy, contributes the most to the overall alignment signal of satellites and centrals at low redshift. There is a transition from a negative $\eta(r)$ at small separations to a positive $\eta(r)$ at separations above $1-2$ Mpc. Although the satellite alignment around the central galaxies is not prominent at low redshift, there is a tangential alignment present at $z=2$ (top right panel).

The bottom row of Figure \ref{fig:satcos2} shows the satellite-satellite alignment, which has a transition from radial to tangential alignments from low to high redshift (left to right panel). This is evidenced for both simple and reduced inertia tensor. At high redshift, we have verified that the tangential alignment signal comes from galaxies in low-mass subhaloes ($M_h<10^{12}$ M$_\odot$), while high mass satellites are radially aligned. On the contrary, the DM subhaloes do not exhibit such transition. Their shape alignment is always radial, regardless of mass and redshift.

\subsection{Spin alignments}
\label{sec:spins}

\begin{figure}
  \centering
  \includegraphics[width=0.49\textwidth]{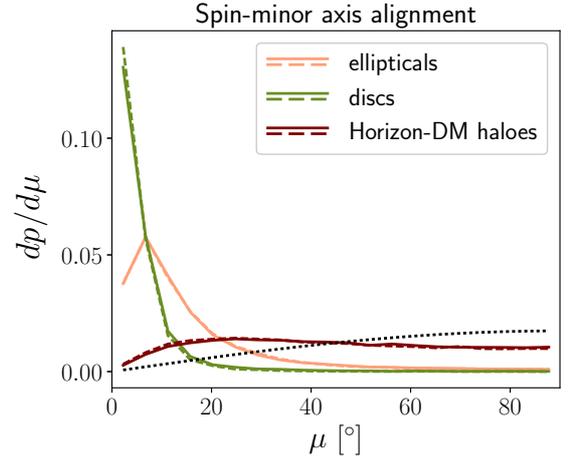}
  \caption{The relative angle between the spin and the minor axis, $\mu$, for ellipticals (orange/light gray), discs (green/dark gray), and matching Horizon-DM haloes (red/black) at $z=0.06$. Results are very similar for simple (solid, $SIT$) and reduced inertia tensors (dashed, $RIT$). The dotted curve represents the expectation for random relative orientations.}
  \label{fig:spin_minor}
\end{figure}

We have focused on prescribing the orientation of a galaxy using its minor axis and comparing it with the halo minor axis. Alignment models for disc galaxies often rely on the relation between the {\it spin} axis of galaxies and haloes instead. We compare the orientation of galaxy and halo minor axes with their respective spin axes. This is shown in Figure \ref{fig:spin_minor} for both reduced (dashed) and simple (solid) inertia tensor. This figure shows there is a strong alignment between the minor and the spin axes of discs (green/dark gray), with a mean misalignment of $4.9^\circ \pm 5.3^\circ$ for the reduced inertia tensor, and a similar value for the simple inertia tensor. Ellipticals (orange/light gray) show a misalignment that is three times larger, also with larger dispersion, as expected for pressure supported systems. Matched haloes in Horizon-DM (red/black) have very little correlation between the direction of their spin and minor axes, with typical misalignments of $\sim 45^\circ$ and negligible difference between disc or elliptical hosts.

\begin{figure}
  \centering
  \includegraphics[width=0.49\textwidth]{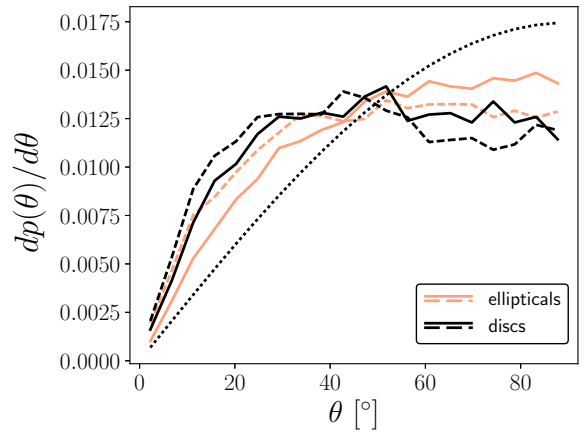}
  \caption{The relative angle between the spin of a galaxy and the matching halo in Horizon-DM (solid) compared to the angle between the minor axes of galaxy and halo from the reduced inertia tensor shape (dashed) at $z=0.06$. Elliptical galaxies are identified in orange/gray, while the relations for discs are shown in black.}
  \label{fig:spinangle}
\end{figure}

Is the spin of discs better correlated with the spin of the matched halo than it is with its minor axis? Figure \ref{fig:spinangle} shows the misalignment angle between galaxy spin axis and halo spin axis (solid) for different galaxy types, where the halo has been cross-matched from the Horizon-DM simulation. For comparison, we also show the distribution of misalignment angles between the minor axis of the galaxy and the halo, using the reduced inertia tensor (dashed). We find that there is little difference between the distribution of misalignment angles between spins or between minor axis, with a slight trend for alignments to be stronger for minor axes, even in the case of discs. This suggests there is no particular advantage to modelling intrinsic alignments of discs by relating their orientation to that of the halo spin axis. Our results are very similar at higher redshifts.

\begin{figure}
  \centering
  \includegraphics[width=0.49\textwidth]{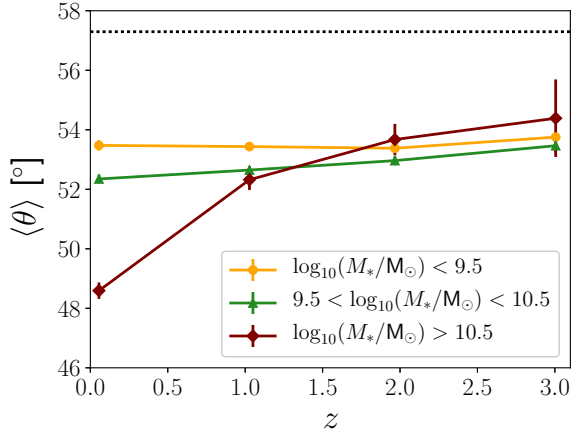}
  \caption{The mean misalignment angle between galaxy and halo spin from Horizon-DM as a function of redshift for galaxies in different mass bins. The dotted line indicates the random expectation.}
  \label{fig:spinanglez}
\end{figure}

Figure \ref{fig:spinanglez} shows the redshift evolution of the mean misalignment angle between the spin of the galaxy and the spin of the matched Horizon-DM halo. There is no evolution for galaxies in low and intermediate mass bins. For high mass galaxies, the trend is similar to that of the middle panel of Figure \ref{fig:misevol} for shapes, with slightly increased misalignment in this case.

\begin{figure}
  \centering
  \includegraphics[width=0.48\textwidth]{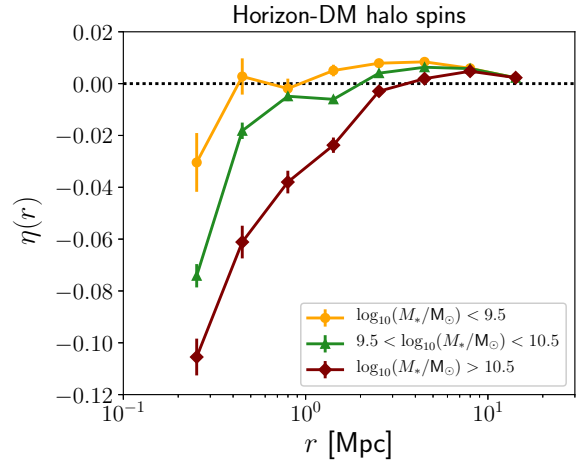}
  \caption{Position-orientation correlation, $\eta(r)$ (Equation \ref{eq:eta}), for halo spins in different mass bins (satellites are included). Diamonds represent galaxies with $M>10^{10.5}{\rm M}_\odot$, triangles correspond to galaxies with $10^{9.5}{\rm M}_\odot<M<10^{10.5}{\rm M}_\odot$, and filled circles, to galaxies with $M<10^{9.5}{\rm M}_\odot$. A negative signal corresponds to the spin axes being perpendicular to the separation vector.}
  \label{fig:spincorr}
\end{figure}

We also measure spin correlations among haloes in Horizon-DM. The results for $z=0.06$ are shown in Figure \ref{fig:spincorr} (for a comparison with minor axis alignment trends, see Figure \ref{fig:cos2}). We find a significant difference in the behaviour of halo spin alignments, compared to halo shape alignments. At low redshift, there is a transition from a negative $\eta(r)$ at small scales to a positive $\eta(r)$ at large separations. The transition is mass dependent, with higher mass haloes exhibiting this change at larger separations. We have verified that this transition is related to the relative alignment of the spin of a central halo and the location of satellites. Haloes with their spin preferentially aligned perpendicular to their minor axis give rise to the positive signal; while haloes with their spin preferentially parallel to the minor axis give rise to a negative trend. In conjunction, the two populations produce a sign transition in $\eta(r)$. This qualitatively different behaviour between spin and shape correlations of haloes implies that spin correlations from DMO simulations should not be used to predict shape alignments of galaxies.


\subsection{Shapes and alignments of galaxies without matching haloes}
\label{sec:nomatch}

The results presented so far are based on the sample of galaxies in Horizon-AGN which has been matched to a host halo, and cross-matched to a halo in Horizon-DM. As was shown in Figure \ref{fig:matcheff}, a significant fraction of galaxies do not have matching haloes in the two Horizon runs. In this section, we explore whether the shapes and orientations of these ``unmatched'' galaxies differ from those in our fiducial sample of centrals and satellites. The number of galaxies in each mass bin for the unmatched sample is $10,010$ for $M\leq 10^{9.5}$ M$_\odot$, $8,069$ for $10^{9.5}$ M$_\odot$ $< M \leq 10^{10.5}$ and $1,408$ for M$_\odot$ and $M> 10^{10.5}$ M$_\odot$ at $z=0.06$. The fraction of ellipticals in each bin is considerably higher than for the matched sample: $\{98,80,47\}\%$ compared to $\{94,47,37\}\%$ for the fiducial sample, from low to high mass, respectively. 

\begin{figure*}
  \centering
  \includegraphics[width=0.85\textwidth]{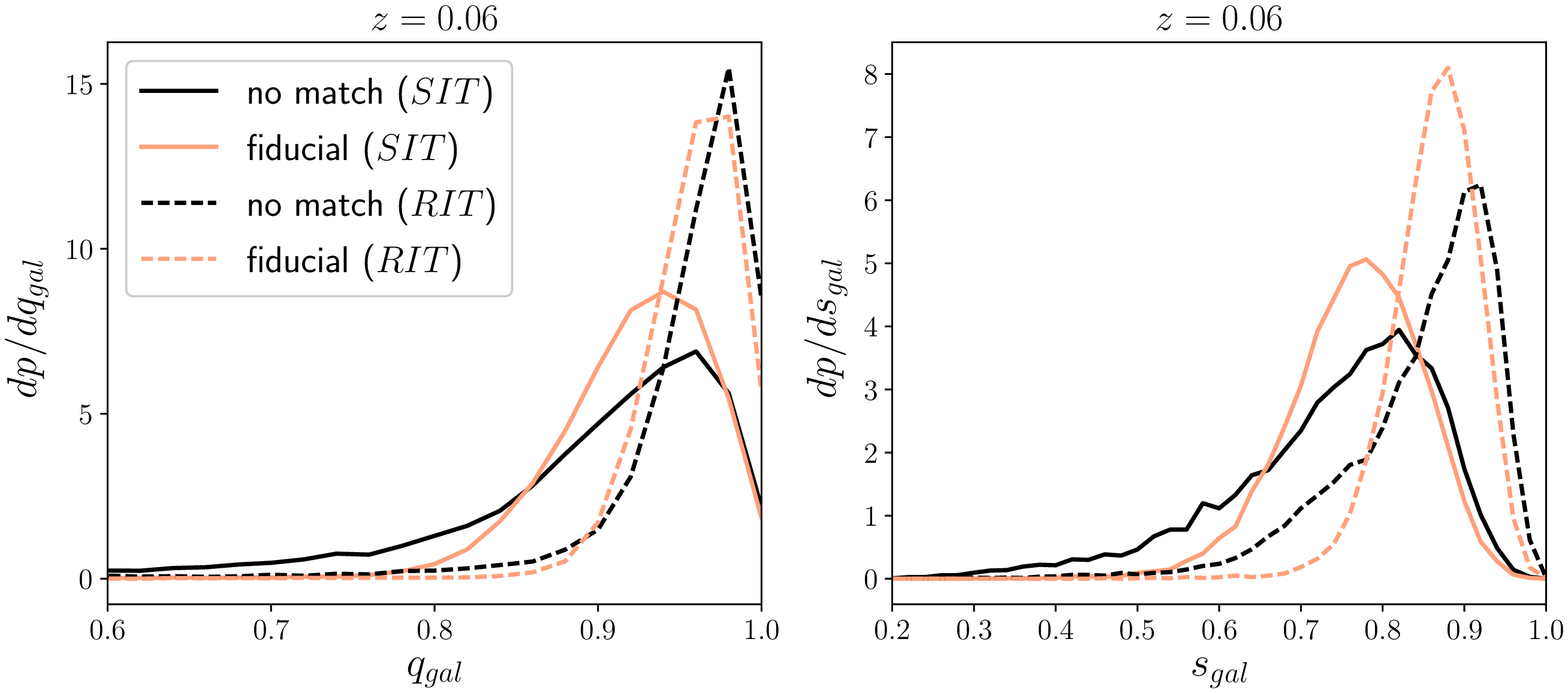}
  \includegraphics[width=0.85\textwidth]{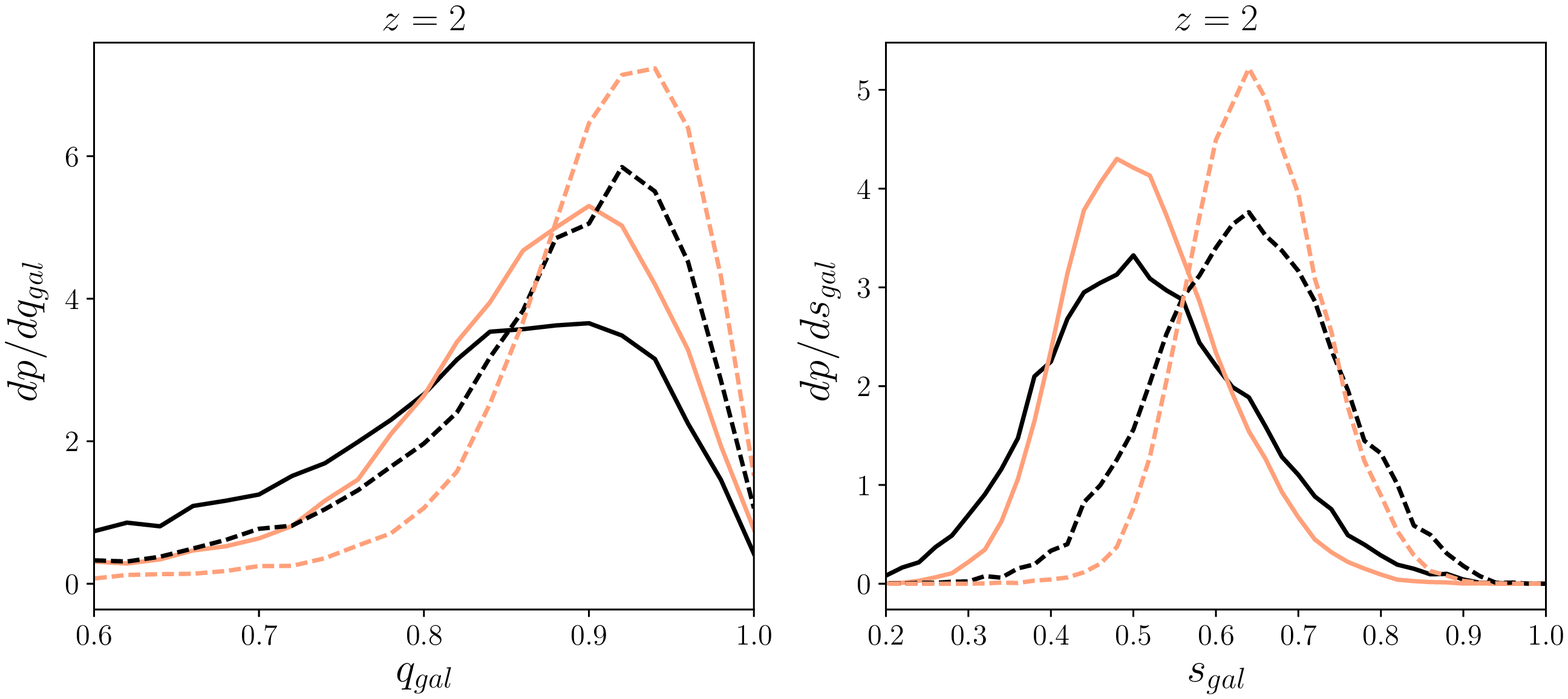}
  \caption{Distribution of axis ratios of galaxies with (gray) and without (black) a match to Horizon-DM haloes. The top panels correspond to $z=0.06$ and the bottom panels, to $z=2$. The left panel indicates the distribution of $q_{gal}$ values, and the right panel, that of $s_{gal}$ values. The dashed curves correspond to reduced inertia tensor measurements and the solid curves, to simple inertia tensor measurements. Weights have been applied to account for the different mass functions of the two samples following Figure \ref{fig:matcheff}. However, results are very similar without applying weights.}
  \label{fig:nomatch_axratio}
\end{figure*}

We consider whether unmatched galaxies have different shape distribution than the fiducial sample. Figure \ref{fig:nomatch_axratio} presents the distributions of axis ratios $q$ and $s$ for both samples at $z=0.06$. We find that there are significant discrepancies in the case of the simple inertia tensor. For the reduced inertia tensor, shapes tend to be rounder, as evidenced by the distributions shifting to values closer to $1$, and the discrepancies between the two populations are reduced. Nevertheless, failed matches are in general more flattened. We have applied a weight to the results in Figure \ref{fig:nomatch_axratio} that compensates for the different mass functions of the two samples. We have found our results to be unaffected by this precaution.

\begin{figure*}
  \centering
  \includegraphics[width=0.85\textwidth]{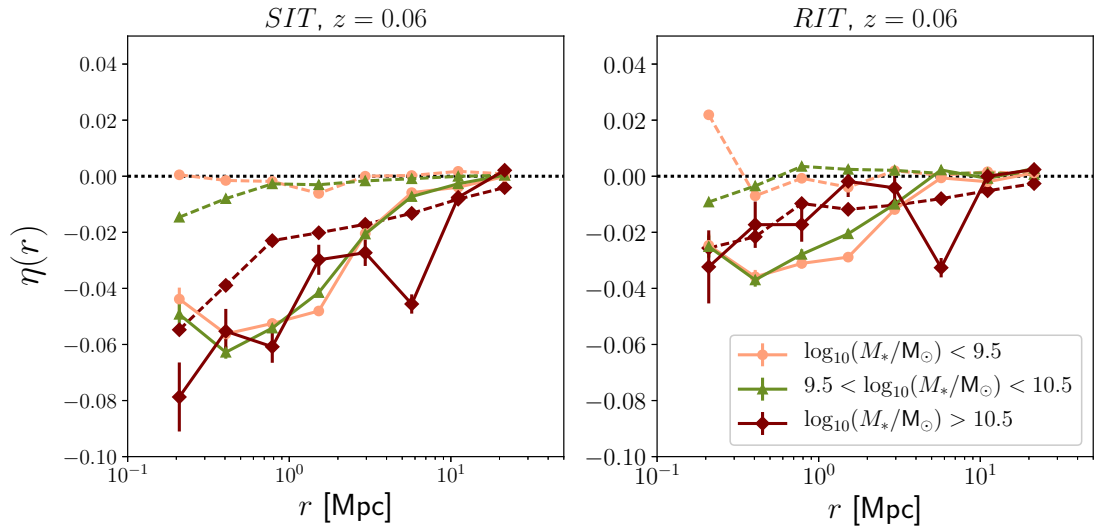}
  \caption{Position-orientation correlation, $\eta(r)$, for failed matches at $z=0.06$ for simple (left panel) and reduced (right panel) inertia tensors. The dashed curves, shown for comparison, correspond to the measurements obtained from the fiducial cross-matched sample.}
  \label{fig:nomatch_corr}
\end{figure*}

Figure \ref{fig:nomatch_corr} shows the $\eta(r)$ statistic for unmatched galaxies in our fiducial mass bins at $z=0.06$, compared to the full fiducial sample, including both centrals and satellites. We find that failed matches at $z=0.06$ have stronger alignment correlations than the fiducial sample. This could be a consequence of inefficient matching for more elongated galaxies and haloes. In other words, distortion of the galaxy and halo shape by the action of the tidal field could reduce the efficiency of the matching. Overall, the results from Figure \ref{fig:nomatch_axratio} and Figure \ref{fig:nomatch_corr} suggest that the shapes and orientations of the failed matches should be modelled differently than those of the fiducial sample. We have verified that their overall contribution to the total intrinsic alignment $\eta(r)$ for all galaxies is significant, and should be accounted for.

\section{Discussion}
\label{sec:discuss}

\subsection{General implications for  models of alignment}

In this work, we have related the shapes and orientations of galaxies to those of their host dark matter haloes in the Horizon-AGN simulation, and to the matching haloes in the Horizon-DM simulation. Our analysis was performed in three dimensions to profit from the full information contained in the simulations. Typically, one would expect to construct a halo model of alignments by relating the galaxy shape and orientation to halo shape and orientation. By comparing galaxies in Horizon-AGN to haloes in Horizon-DM, we have found that the shape of a galaxy is poorly related to the shape of the halo. A better option would be to model galaxy shapes as drawn from distributions of shapes for discs and ellipticals at a given redshift. We have characterised the misalignment angle between the minor axis of galaxies and haloes, finding that this angle is a strong function of halo mass, with significant residuals in the case of high redshift ellipticals. This suggests that the misalignment angle should be modelled independently for galaxies of different types. We have also found no advantage in using halo spin as a proxy for disc orientation.

Our results also have implications for intrinsic alignment modelling, particularly for the redshift evolution of alignments. We have found that there is a striking difference between the alignment of galaxies and those of haloes. Halo shape alignments are radial towards each other and they decrease in amplitude with time. The alignments of galaxies, on the other hand, are influenced by the fact that galaxies tend to align better with the host halo over time. In terms of the tidal alignment model \citep{Catelan01,Hirata04}, this suggests that the ``alignment bias'' of galaxies and haloes with respect to the tidal field of the large-scale structure has a different redshift evolution. (Notice that the use of cross-matched samples of the simulations implies that the galaxies and haloes, as we have selected them, have the same clustering bias.)

 The effect of baryonic feedback on halo alignment is significant, resulting in a fractional decrease of the $\eta(r)$ statistic of $10-30\%$ when comparing the Horizon-DM to Horizon-AGN halo alignments. As expected, the effect increases towards low redshift. This implies that a successful recipe to populate DMO simulations with alignments must directly map the galaxy shape and orientation to that of the Horizon-DM halo. In other words, assuming that haloes in the DMO simulation are unaffected by baryonic feedback can lead to biases in the predicted intrinsic alignment signal. 

Since the pioneering work by \citet{Sastry68}, \citet{Holmberg69} and \citet{Binggeli82}, many observational works in the literature have confirmed that satellites are distributed anisotropically around their host \citep{Brainerd05,Azzaro07,Agustsson10,Niederste10,Hao11}. In addition, satellites could be preferentially oriented around the central host \citep{Pereira05,Faltenbacher07}, although other works have not been able to confirm this trend \citep{Hao11,Schneider13,Chisari14,Sifon15}. To explore these alignment mechanisms, we have selected a satellite sample of galaxies by identifying galaxies matched to Horizon-DM subhaloes. This selection ensures that no information is needed from Horizon-AGN to classify galaxies as satellites; only the Horizon-DM information is used. We find that satellites have similar misalignment PDFs as centrals and they contribute significantly to the intrinsic alignment signal at the scales probed in this work (see Figure \ref{fig:cos2}). As presented in Section \ref{sec:satellites}, the contributions from satellite alignments arise from the preferential orientation of satellite galaxies around centrals, and from the anisotropic distribution of satellites in the direction of the major axis of the central. We had found similar trends in \citet{Chisari16} for galaxies applying mass selection; although we did not explicitly identify a sample of satellites in that work. 

We have seen in Section \ref{sec:corr} that the distribution and orientation of satellite galaxies with respect to centrals has different qualitative behaviour at low and high redshift. At high redshift, satellites are aligned tangentially around centrals, and this alignment becomes negative towards low redshift. This suggests that the signal is dominated by the presence of disc galaxies, with their minor axes pointing in the direction of the filaments that connect centrals. The distribution of satellites is anisotropic, and it is related to the orientation of the central. At $z=2$, satellites are preferentially located in the direction of the minor axis of centrals at large scales. As structure grows, torquing from the central galaxies leads to preferential distribution of satellites along the major axis of the central. This interpretation is consistent with the findings of \citet{Welker15}, who analysed how satellites settle into a plane around their central galaxy in Horizon-AGN.

Notice that our matching procedure, which identifies central galaxies as the most massive galaxy within the sphere of $10\%$ $R_{\rm vir}$ of a certain halo preferentially discards satellites within this volume unless matched to a different subhalo. As a result, the population of ``failed matches'' (Section \ref{sec:nomatch}) could have a significant fraction of satellites which are not captured in our fiducial analysis. As mentioned in Section \ref{sec:nomatch}, additional prescriptions would need to be developed to model the intrinsic alignments of this population.


\subsection{Halo models of alignment}

Halo models of alignments \citep{Heymans04, Schneider10,Joachimi13b} are built upon several assumptions which we have put to the test in this work. 

\citet{Heymans04} relied on the direction of the angular momentum of dark matter haloes to estimate the intrinsic alignment signal of galaxies. They connected the orientation of a galaxy to the orientation of the halo angular momentum by imposing a Gaussian distribution for the misalignment angle with non-zero mean derived from a relation between the angular momentum of the gas and that of the halo obtained found in simulations by \citet{Bosch02}. As shown in Section \ref{sec:spins}, significant biases in predicting the alignment signal can arise from such a model, as the alignments between the spins of haloes throughout the large-scale structure show qualitatively different behaviour to the alignment of galaxy shapes. 

\citet{Joachimi13b} split the galaxy population into ellipticals and discs. For ellipticals, they use the reduced inertia tensor of the halo as a proxy for the shape. We have shown in Section \ref{sec:shapes} that these bear little correlation, particularly to the shape of matched haloes in DMO simulations and under identical initial conditions. In that work, the relative misalignment between the projected major axis of an elliptical and its host halo was assumed to follow a Gaussian distribution with zero mean and a scatter of $35^\circ$ \citep{Okumura09}. We show in appendix \ref{app:projected} that this can result in an overestimate of the alignments of ellipticals, as there is significance mass dependence in the distribution of the misalignment angle.

For discs, \citet{Joachimi13b} define the orientation of the minor axis to coincide with the spin axis of the host halo. We have found that the halo spin is less well-aligned with the orientation of a disc than the halo minor axis. Moreover, in general, galaxy orientations assigned by adopting halo orientations will overestimate the intrinsic alignment signal, and will not capture the correct redshift dependence of alignments, as seen in sections \ref{sec:corr} and \ref{sec:spins}. At the very least, redshift dependent misalignment angle distributions should be used, and these are a function of other galaxy properties (such as galaxy dynamics). We have not found any evidence for environmental dependence other than that caused by halo mass variations and by the distinction between centrals and satellites.

\citet{Schneider10} apply their model to ellipticals alone. For centrals, they consider the predictions given by the linear alignment model \citep{Catelan01,Hirata04,Hirata10}, without relating the shape and orientation of the galaxy to that of the host halo. This is accurate by construction if observational constraints on the linear alignment model are considered. Horizon-AGN alignments have also been shown to have a scale and redshift dependence consistent with the linear alignment model at large scales \citep{Chisari15,Chisari16}. Satellite galaxies in their model are isotropically distributed around the center of the halo and point radially towards it. The relative misalignment angle between the satellite major axis and the separation vector follows a distribution extracted from DMO simulations. Based on the results of this work, potential improvements that need to be considered are the {\it anisotropic} distribution of satellites (as shown in section \ref{sec:satellites}; see also \citealt{Welker15}) and the relative misalignment between satellite galaxies and their host.    

\subsection{Comparison to other hydrodynamical simulations}

Other works have explored the relation between galaxy and host halo alignment in other cosmological numerical hydrodynamical simulations. \citet{Tenneti14a} measured the shapes of galaxies and haloes in the MassiveBlack-II simulation, a smoothed-particle-hydrodynamics simulation with baryonic feedback implementation. They restricted their sample to galaxies and haloes with more than $1000$ stellar and dark matter particles. The lower stellar mass probed in that work was $10^{10}\,h^{-1}\,{\rm M}_\odot$, which corresponds to our intermediate mass range. Hence, we shall compare our results above this threshold only. \citet{Tenneti14a} measured galaxy and halo shapes using the simple inertia tensor (Eq. \ref{eq:sit}) and found that dark matter haloes are typically rounder than stellar haloes, with axis ratios decreasing towards higher masses. Similarly to our Fig. \ref{fig:axesratio}, they find a dip in the mean axes ratio $\langle s \rangle$ of galaxies at intermediate masses ($M\sim 10^{11}{\rm M}_\odot$). We attribute this to the fact that discs dominate the galaxy population in this mass range, driving $\langle s \rangle$ to lower values than the value found for ellipticals. This can be explicitly seen in the elliptical vs. disc decomposition of the curves in Figure \ref{fig:axesratio}.

Regarding galaxy-halo misalignment, they find, in agreement with our results, that the mean misalignment angle decreases towards high masses. However, they focus on measuring the misalignment angle of the {\it major} axis, rather than the minor axis. The mean misalignment angle between galaxies and massive halo hosts ($M_h>10^{13}\,h^{-1}\,{\rm M}_\odot$) was $\sim 10^\circ$, while we find a $\sim 29^\circ$ misalignment for the same population in Horizon-AGN. The minor axis is better aligned in Horizon-AGN, at a comparable level to the degree of misalignment found for the {\it major} axis in MassiveBlack-II. \citet{Tenneti14a} find no evidence of redshift evolution of the mean misalignment at fixed mass, contrary to our results from Figure \ref{fig:misevol}. 

\citet{Tenneti15c} compared galaxy-halo alignments in MassiveBlack-II to a twin DMO run. They restricted to haloes with over $90\%$ matching efficiency between the two simulations, resulting in an effective mass threshold of $M_h>10^{10.8}\,h^{-1}\,{\rm M}_\odot$. This is a slightly larger mass than the least massive haloes considered in this work. Their shape measurements are based on an iterative calculation of the reduced inertia tensor. They found that haloes are rounder when baryons are present, consistently with the results presented in Figure \ref{fig:axesratio}. 

In agreement with our findings, low mass galaxies in \citet{Tenneti15c} are more misaligned with their embedding haloes than high mass galaxies. The alignment is stronger with the central region of haloes, rather than with the outskirts, as was seen in Figure \ref{fig:align}. Also, the alignment of the galaxy with the halo in the DMO simulation is less sensitive to whether one is probing the inner or outer regions of the halo. We take a step further in Section \ref{sec:resid} by showing that the mean galaxy-halo alignment is not only a function of mass, but dynamical properties of the galaxies might be playing a role in determining this alignment. We also test whether the mean misalignment angle is correlated with the large-scale structure, but found no environmental dependence beyond mass. \citet{Tenneti15c} also computed the $\eta(r)$ correlation between galaxies, and between haloes. They found, like we do in Figure \ref{fig:cos2}, that galaxy alignment is suppressed with respect to halo alignment. 

\citet{Velliscig15} studied galaxy-halo alignments in the EAGLE and Cosmo-OWLS simulations. Similarly to MassiveBlack-II, these simulations adopt a smoothed-particle-hydrodynamics technique, although the implementation of baryonic feedback processes differs between them. \citet{Velliscig15} found that early-type galaxies are more misaligned with their host haloes compared to late-type galaxies. However, this has been shown to be sensitive to the particular criterion used for distinguishing discs and ellipticals. When adopting a distinction based on the degree of ordered rotation \citep{Scannapieco09}, \citet{Shao16} found that ellipticals living in Milky Way-mass haloes in the EAGLE simulation have better alignment with their haloes than disc galaxies. In Horizon-AGN, we find similar average misalignments for both populations of discs and ellipticals at $z=0.06$, but high mass ellipticals display a clear trend for small misalignment angles, as they tend to populate high mass haloes (see Figure \ref{fig:avgmis}). \citet{Velliscig15} provide misalignment angle distributions for different halo masses to be used for populating DMO simulations. Our results, and those of \citet{Shao16}, suggest that distinguishing between late- and early-type is necessary to obtain accurate predictions of misalignment angles (Figure \ref{fig:resid}). On the other hand, \citet{Velliscig15b} studied the correlations between galaxy orientations in EAGLE at $z=0.06$ and found similar qualitative trends for the alignments of satellites as found in this work. In addition, they remarked that using halo orientation as proxy for galaxy orientation produces an overestimate of the intrinsic alignment signal.

The main point of disagreement between these works and our results from Horizon-AGN is that, as remarked in \citet{Tenneti15b}, the alignments of disc galaxies have opposite signs, i.e., discs align tangentially around overdensities of the matter field in Horizon-AGN \citep{Chisari15,Chisari16}. This prediction for tangential alignment is in agreement with expectations from tidal torque theory \citep{Codis15}. In addition, we stress that Horizon-AGN and MassiveBlack-II predict a different redshift dependence of the misalignment angle between galaxies and haloes. Halo models of alignments should take into account these discrepancies by allowing for sufficient flexibility until these predictions can be further tested by current weak lensing surveys.

\section{Conclusions}
\label{sec:conclusion}

In this manuscript, we have studied the galaxy-halo alignment in the Horizon-AGN cosmological hydrodynamical simulation and in Horizon-DM, a twin DMO run with the same initial conditions. In particular, we focused on understanding the relation between galaxy shape and orientation and halo shape and orientation. This allows us to reach the following conclusions, which are relevant for building halo models of intrinsic alignments:
\begin{itemize}
\item The shape of a galaxy cannot be related to that of the matching halo in a DMO simulation. Nevertheless, galaxy shapes are a function of halo mass, but discs and ellipticals should be modelled separately, since they undergo different mass and angular momenta assembly histories.
\item Similarly, the misalignment angle between the minor axis of a galaxy and the matching halo is a strong function of halo mass, but there are significant residuals from this relation when the galaxy population is split into discs and ellipticals. In addition, we find a significant dependence of the misalignment angle between the minor axis of galaxy and halo on the distance to the nearest filament; however, this arises due to dependence of the misalignment angle on halo mass.
\item Haloes in both Horizon-AGN and Horizon-DM align their minor axis perpendicularly towards each other. They decrease their shape alignment with each other towards low redshift, and the alignment is weaker if the halo hosts a low mass galaxy. Halo alignments are stronger in Horizon-DM, suggesting that the impact of baryons is to decrease the alignment with the large-scale structure at fixed mass cuts. In fact, baryons cause a mass-dependent relative misalignment of haloes between the Horizon-DM and Horizon-AGN simulation, which is more prominent at high mass.
\item Galaxy shape alignments, on the other hand, are much weaker, overall, than halo alignments. Although halo alignments are decreasing towards low redshifts, galaxies tend to compete with this effect by ``catching-up'' with their host, increasing their relative alignment with the host halo. We hypothesize that this is possibly due to the colder component taking longer to settle. Indeed, disc orientations are set up by the gas flows along the connecting large-scale structure,  which are steady throughout cosmic history \citep{Pichon11,Stewart13}. On the other hand, the dark halo's orientation is also set up by the same large scale tide but, being dynamically hotter, it reacts less to the last small merger events; it will therefore settle first.
\item We identified satellite galaxies as those matched to substructure in dark matter haloes. These satellites exhibit an anisotropic distribution about the minor axis of the central galaxy, which should be considered in halo models of alignment. In addition, they also display preferential orientations with respect to the central galaxy, being subject to a transition from radial to tangential alignments of their minor axis from high to low redshift. 
\item We have explored whether the spin of haloes provides additional information for inferring galaxy alignment. We have found that even in the case of discs, spins do not seem to present any advantage to the modelling. Moreover, the spins of haloes display significantly different correlations than their shapes, implying that caution must be taken to relate spins to shapes in models of intrinsic alignment.
\end{itemize}

Overall, we have tested in this work several of the assumptions of current halo models of alignments and established areas for improvement in how to connect galaxy shapes and orientations to those of the cross-matched haloes from DMO simulations. Our conclusions in this section provide a road-map for connecting galaxy alignments with DMO outputs. One important caveat is that the sample of ``unmatched'' galaxies remains a significant contributor to the overall alignment signal. As a consequence, we envisage that future work should focus on developing a scheme to account for the missing galaxies.

\section*{Acknowledgements}
This work has made use  of the HPC resources of CINES (Jade and Occigen supercomputer) under the time allocations 2013047012, 2014047012 and 2015047012 made by GENCI. This work is partially supported by the Spin(e) grants {ANR-13-BS05-0005} (\url{http://cosmicorigin.org}) of the French {\sl Agence Nationale de la Recherche} and by the ILP LABEX (under reference ANR-10-LABX-63 and ANR-11-IDEX-0004-02). Part of the analysis of the simulation was performed on the DiRAC facility jointly funded by the Science and Technology Facilities Council (STFC), BIS and the University of Oxford. NEC acknowledges support from a Beecroft fellowship. The work of NK was supported by the Oxford Astrophysics Summer Research Programme. RSB is funded by the STFC. LM is supported by STFC grant ST/N000919/1. We thank S. Rouberol for smoothly running the {\tt Horizon} cluster for us.  We are grateful to David Alonso for his comments on the impact of grid-locking. We acknowledge useful discussions with Risa Wechsler and Renyue Cen. 

\bibliographystyle{mn2e_warx}
\bibliography{misalign}

\appendix

\section{Impact of grid-locking}
\label{app:gridlock}

\begin{figure}
  \centering
  \includegraphics[width=0.47\textwidth]{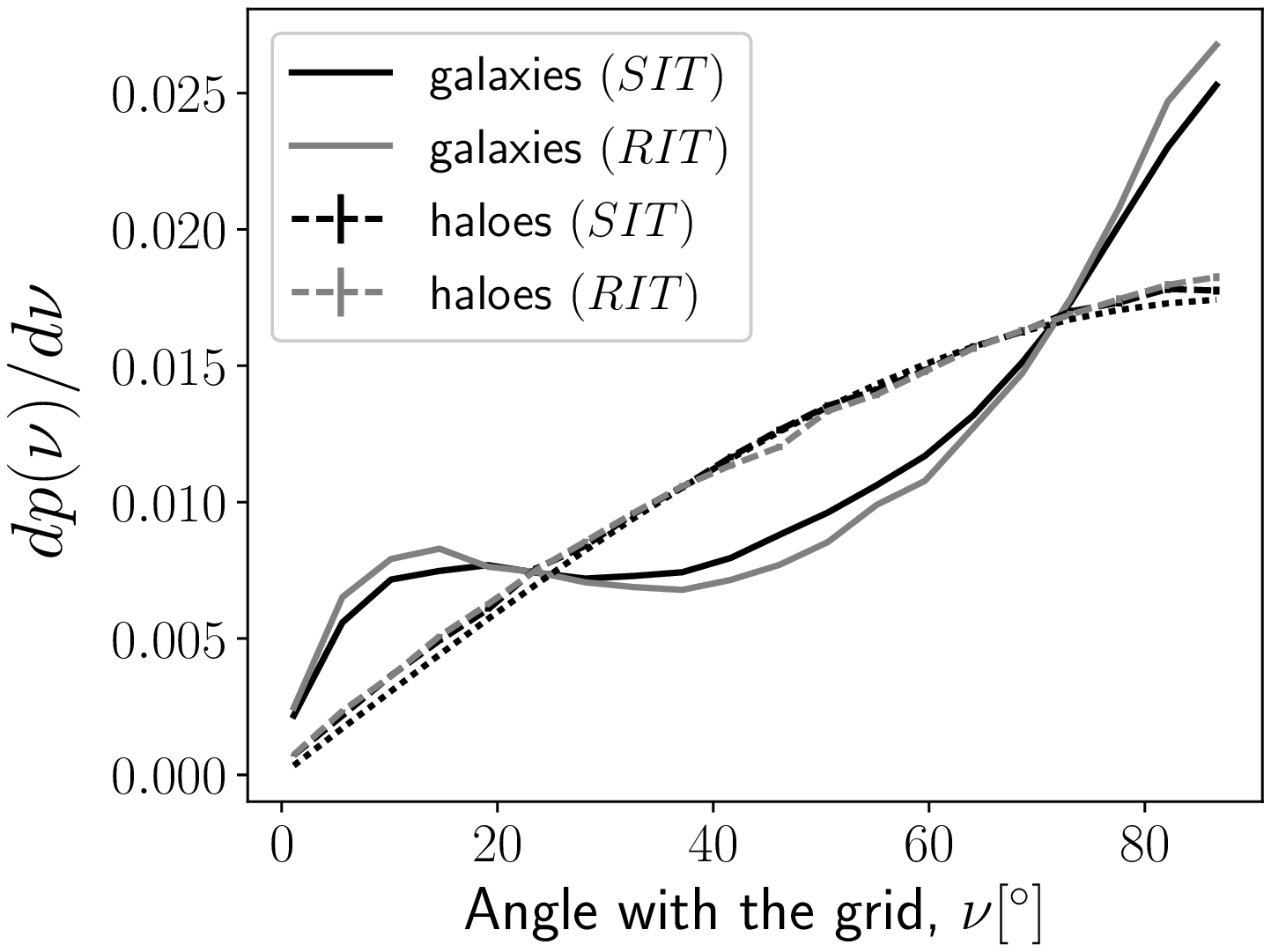}
  \caption{PDF of the angle between a galaxy (solid) or a dark matter halo (dashed) with the grid at $z=0.06$. The behaviour is similar for $\{x,y,z\}$, so here we have averaged over the three directions. The dotted line represents the expectation from a random distribution of orientations of the eigenvectors.}
  \label{fig:gridxyz}
\end{figure}

Grid-locking refers to the preferential alignment of the spins and shapes of galaxies with the $\{x,y,z\}$ directions of the grid as a consequence of numerical artifacts in AMR simulations. This correlation is explicitly shown in Figure \ref{fig:gridxyz} for galaxies (solid curves) and dark matter haloes (dashed curves) and compared to a random distribution (dotted) at $z=0.06$. The galaxies are not randomly oriented with respect with the grid; there is an excess of galaxies at the directions parallel to one axis or perpendicular to it (parallel to the other two axes). This is not the case for dark matter haloes, whose minor axis vectors are uncorrelated with the box. The mean misalignment angle between haloes and the grid is consistent with random at $<2\sigma$.

In view of the periodic boundary conditions of the box, we expect that, by symmetry, the grid-locking component of the shape of a galaxy will average to null around any arbitrary point within it. In \citet{Chisari15}, we directly tested this hypothesis by computing the orientation of galaxy shapes and spins around random points in the simulation box. We successfully showed that this results in a null signal. As a consequence, the $\eta$ statistic defined in Eq. (\ref{eq:eta}) is not affected by grid-locking. 

Misalignment angle distributions for galaxies with respect to dark matter haloes are not biased by grid-locking either. While individual misalignment angles can be biased, the average over a significant number of galaxy-halo pairs should cancel this bias by symmetry. This is because halo orientations are not grid-locked, as shown in Figure \ref{fig:gridxyz}. Halo orientations are effectively random vectors inside the box, making this case completely analogous to the example described in the previous paragraph. Hence, misalignment angle distributions and average misalignment angles are not grid-locked. Nevertheless, grid-locking can contribute to worsening the noise properties of the misalignment statistics. The effect of grid-locking decreases significantly towards higher redshift.

\section{Projected misalignment}
\label{app:projected}

The main body of this manuscript focuses on describing the relation between the three-dimensional orientation of a galaxy and that of the matching halo, as described by the minor axis of the ellipsoid corresponding to the inertia tensor. Other works phrase this relation in terms of the relative orientation between the {\it projected} shapes, rather than using the full three-dimensional information. The motivation for this alternative description is that observations give us access to projected shapes only. In this work, we have preferred to provide misalignment measurements in three dimensions, to extract as much information as possible from the hydrodynamical simulations. For comparison with the results of \citet{Okumura09} for luminous red galaxies, which were used in the model of \citet{Joachimi13b}, we provide in this appendix misalignment angle distributions for the {\it major} axis of the projected shapes of elliptical galaxies.

\begin{figure}
  \centering
  \includegraphics[width=0.47\textwidth]{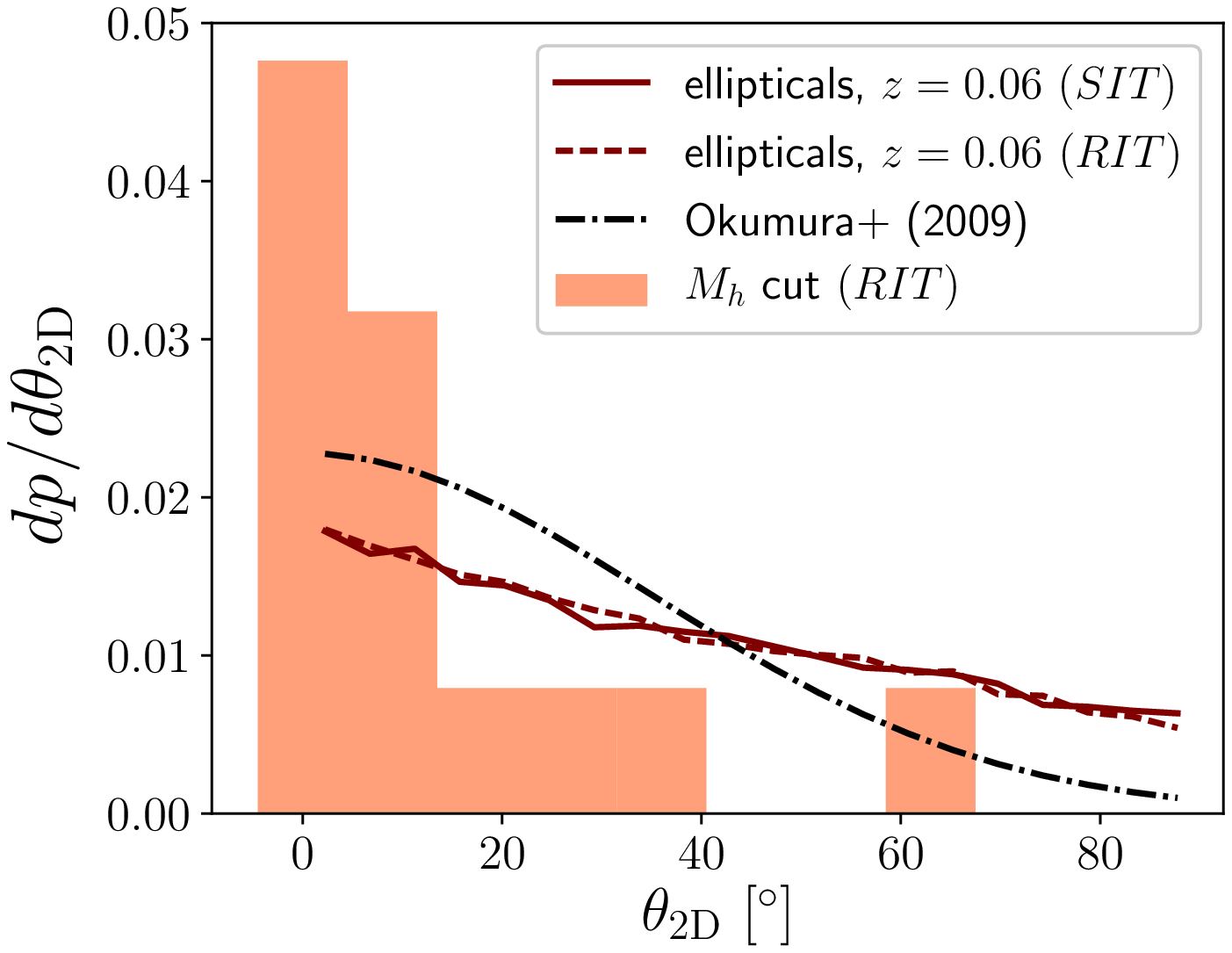}
  \caption{PDF of the angle between the major axis of the projected shape of a galaxy and its matched dark matter halo in Horizon-DM at $z=0.06$. The curves for elliptical galaxies are the solid curve, corresponding simple inertia tensor, and the dashed line, for the reduced inertia tensor of both galaxy and halo. The dot-dashed line is the distribution proposed by \citet{Okumura09}. The shaded areas represent the distribution of misalignment angles for haloes above the mass cut of the halo model by \citet{Zheng09} for luminous red galaxies.}
  \label{fig:major2D}
\end{figure}

We obtain projected shapes for both elliptical galaxies and matched haloes in Horizon-DM by reducing the inertia tensor calculation (Eqs. \ref{eq:sit} and \ref{eq:rit}) to two dimensions, excluding one of the coordinates of the cosmological box. The eigenvector of the inertia tensor associated to the largest eigenvalue indicates the direction of the major axis. We compare the direction of the major axes of galaxies and matched haloes in Horizon-DM and obtain the misalignment angle distribution shown in Figure \ref{fig:major2D}. The red curves correspond to all ellipticals at $z=0.06$, and we show results for the simple (solid) and reduced (dashed) inertia tensors. The black dot-dashed line indicates the distribution proposed by \citet{Okumura09}, a Gaussian function of zero mean and $35^\circ$ scatter. This overestimates the alignments of the overall population of ellipticals, whose error bars are too small to be shown in this figure. Hence, this suggests that applying the proposed \citet{Okumura09} distribution to all ellipticals, as in \citet{Joachimi13b}, could result in an overestimate of the alignment signal.

On the other hand, if we apply a cut on the halo mass corresponding to luminous red galaxies \citep{Zheng09}, the distribution of misalignment angles, shown in orange, has a mean misalignment angle and dispersion of $-2^\circ\pm 48^\circ$ for the simple inertia tensor case, and $6^\circ \pm 23^\circ$ for the reduced inertia tensor, in better agreement with \citet{Okumura09}.

\end{document}